\newif\ifhasbib

\hasbibtrue
\documentclass[twocolumn]{aastex61}

\shorttitle{Exoplanet Population Observation Simulator}
\shortauthors{Mulders et al.}

\usepackage{natbib,graphicx,xspace,amsmath,footnote}
\newcommand{\fdet}{\ensuremath{f_{\rm det}}\xspace}
\newcommand{\fgeo}{\ensuremath{f_{\rm geo}}\xspace}
\newcommand{\fsnr}{\ensuremath{f_{\rm S/N}}\xspace}
\newcommand{\fvet}{\ensuremath{f_{\rm vet}}\xspace}
\newcommand{\fpl}{\ensuremath{f_{\rm pl}}\xspace}
\newcommand{\ID}{{\rm ID}\xspace}

\newcommand{\fiso}{\ensuremath{f_{\rm iso}}\xspace}

\newcommand{\epos}{\texttt{EPOS}\xspace}
\newcommand{\kepler}{\textit{Kepler}\xspace}
\newcommand{\emcee}{\texttt{emcee}\xspace}
\newcommand{\python}{\texttt{Python}\xspace}

\begin{document}

\title{The Exoplanet Population Observation Simulator. I - The Inner Edges of Planetary Systems}

\correspondingauthor{Gijs D. Mulders}
\email{mulders@lpl.arizona.edu}

\author{Gijs D. Mulders}
\affil{Lunar and Planetary Laboratory, The University of Arizona, Tucson, AZ 85721, USA}
\affil{Earths in Other Solar Systems Team, NASA Nexus for Exoplanet System Science}

\author{Ilaria Pascucci}
\affil{Lunar and Planetary Laboratory, The University of Arizona, Tucson, AZ 85721, USA}
\affil{Max Planck Institute for Astronomy, K\''onigstuhl 17, Heidelberg, D-69117, Germany}
\affil{Earths in Other Solar Systems Team, NASA Nexus for Exoplanet System Science}

\author{D\'aniel Apai}
\affil{Department of Astronomy, The University of Arizona, Tucson, AZ 85721, USA}
\affil{Lunar and Planetary Laboratory, The University of Arizona, Tucson, AZ 85721, USA}
\affil{Max Planck Institute for Astronomy, K\''onigstuhl 17, Heidelberg, D-69117, Germany}
\affil{Earths in Other Solar Systems Team, NASA Nexus for Exoplanet System Science}

\author{Fred J. Ciesla}
\affil{Department of the Geophysical Sciences, The University of Chicago, 5734 South Ellis Avenue, Chicago, IL 60637}
\affil{Earths in Other Solar Systems Team, NASA Nexus for Exoplanet System Science}

\begin{abstract}
The \textit{Kepler} survey provides a statistical census of planetary systems out to the habitable zone.
Because most planets are non-transiting, orbital architectures are best estimated using 
simulated observations of ensemble populations.
Here, we introduce \epos, the Exoplanet Population Observation Simulator, to estimate the prevalence and orbital architectures of multi-planet systems based on the latest \textit{Kepler} data release, \texttt{DR25}. 
We estimate that at least $42\%$ of sun-like stars have nearly coplanar planetary systems with 7 or more exoplanets.
The fraction of stars with at least one planet within $1$ au could be as high as $100\%$ depending on assumptions about the distribution of single transiting planets.
We estimate an occurrence rate of planets in the habitable zone around sun-like stars of $\eta_\oplus=36\pm14\%$.
The innermost planets in multi-planet systems are clustered around an orbital period of 10 days (0.1 au), reminiscent of the protoplanetary disk inner edge or could be explained by a planet trap at that location.
Only a small fraction of planetary systems have the innermost planet at long orbital periods, with fewer than $\approx 8\%$ and $\approx 3\%$ having no planet interior to the orbit of Mercury and Venus, respectively. 
These results reinforce the view that the solar system is not a typical planetary system, but an outlier among the distribution of known exoplanetary systems.
We predict that at least half of the habitable zone exoplanets are accompanied by (non-transiting) planets at shorter orbital periods, hence knowledge of a close-in exoplanet could be used as a way to optimize the search for Earth-size planets in the Habitable Zone with future direct imaging missions.
\end{abstract}


\section{Introduction}
The \textit{Kepler} exoplanet survey has revolutionized our understanding of planetary systems around other stars \citep[e.g.][]{2017PAPhS.161...38B}. During the first four years of the mission, hereafter \kepler, it discovered thousands of exoplanets and exoplanet candidates, the majority of which are smaller than Neptune and orbit close to their hosts stars. 
Because transit surveys can detect only a small fraction of the exoplanets, a completeness correction is necessary to understand the true population of exoplanets \citep[e.g.][]{2014PNAS..11112647B}.

The planet occurrence rate, defined as the average number of planets per star, is typically found to be of order unity for planets in the orbital period and planet radius range detectabe with \textit{Kepler} \citep{2011ApJ...742...38Y,2013ApJ...766...81F}.
Much progress has been made in recent years in understanding the survey detection efficiency of the \kepler mission \citep[e.g.][]{2015ApJ...810...95C,2016ApJ...828...99C}.
While earth-sized planets in the habitable zone of sun-like stars are just below the detection limits of \textit{Kepler} \citep{2018ApJS..235...38T}, occurrence rates of earth-sized planets in the habitable zone, $\eta_\oplus$, can be estimated by considering smaller M and K dwarf stars \citep{2013ApJ...767...95D,2015ApJ...807...45D} or by extrapolating from shorter orbital periods and/or larger planet radii \citep{2013PNAS..11019273P,2014ApJ...795...64F,2015ApJ...809....8B}.

The exoplanet population discovered by the \kepler mission has been characterized in great detail. Key findings include that planet occurrence rates increase with decreasing planet size \citep{2012ApJS..201...15H} and remains roughly constant for planets smaller than $3 ~R_\oplus$ \citep{2013ApJ...770...69P,2014ApJ...791...10M,2015ApJ...814..130M}. Recently, a gap in the planet radius distribution has been identified around $1.5-2.0~R_\oplus$ \citep{2017AJ....154..109F,2017arXiv171005398V}. The occurrence of sub-Neptunes increases with distance from the host star up to an orbital period of $\sim10$ days, after which it remains roughly constant \citep{2012ApJS..201...15H,2015ApJ...798..112M}, in contrast to planets larger than Neptune whose occurrence increases with orbital period \citep{2013ApJ...778...53D}.

The presence of multiple transiting planets around the same star provides important additional constraints on planetary orbital  architectures. 
The large number of observed systems with multiple transiting planets indicate that planets are preferentially located in systems with small mutual inclinations \citep{2011ApJS..197....8L,2012ApJ...750..112L,2012ApJ...761...92F}. Their orbital spacings follow a peaked distribution without a clear preference for being in orbital resonances \citep{2014ApJ...790..146F}.
However, the majority of planets in multi-planet system are not transiting, and modifying planetary occurrence rates from those based on the observed planetary architectures to account for non-transiting or otherwise undetected planets is not straightforward.  The true multiplicity of planetary systems can only be constrained from the observed multiplicity and making assumptions about their mutual inclination distribution (\citealt{2012AJ....143...94T}, see also \citealt{2016ApJ...821...47B}).
Ensemble populations of systems with six or more planets with mutual inclinations of a few degrees provide a good match to the observed population of multi-planet systems \citep{2011ApJS..197....8L,2012ApJ...761...92F,2014ApJ...790..146F,2016ApJ...816...66B}.
However, these simulations under-predict the number of single transiting systems by a factor 2, indicating that an additional populations of planets must be present that is either intrinsically single or has high mutual inclinations.

The occurrence rates and planetary architectures place strong constraints on planet formation models \citep[e.g.][]{2013ApJ...775...53H,2016ApJ...822...54D}.
However, the quantitative comparison of planet formation and orbital evolution models and the observed exoplanet population has been greatly complicated by observational biases. Therefore, published comparisons resorted to trying to explain the presence of specific planet sub-populations by identifying processes that can give rise to such planets. While these studies are essential, such qualitative comparisons can not make full use of the wealth of information represented by the overall exoplanet population statistics.
To address this limitation we introduce \epos\footnote{\url{https://github.com/GijsMulders/epos}}, the Exoplanet Population Observation Simulator, that provides a \texttt{Python} interface comparing planet population synthesis models to exoplanet survey data, taking into account the detection biases that skew the observed distributions of planet properties. \epos uses a forward modeling approach to simulate observable exoplanet populations and constrain their properties via Markov Chain Monte Carlo simulation using \emcee \citep{2013PASP..125..306F}, and has already been employed in two different studies \citep{2018arXiv180209602K,2018arXiv180300777P}.

\begin{figure*} % Fig 1
	\centering
    \includegraphics[width=0.8\linewidth]{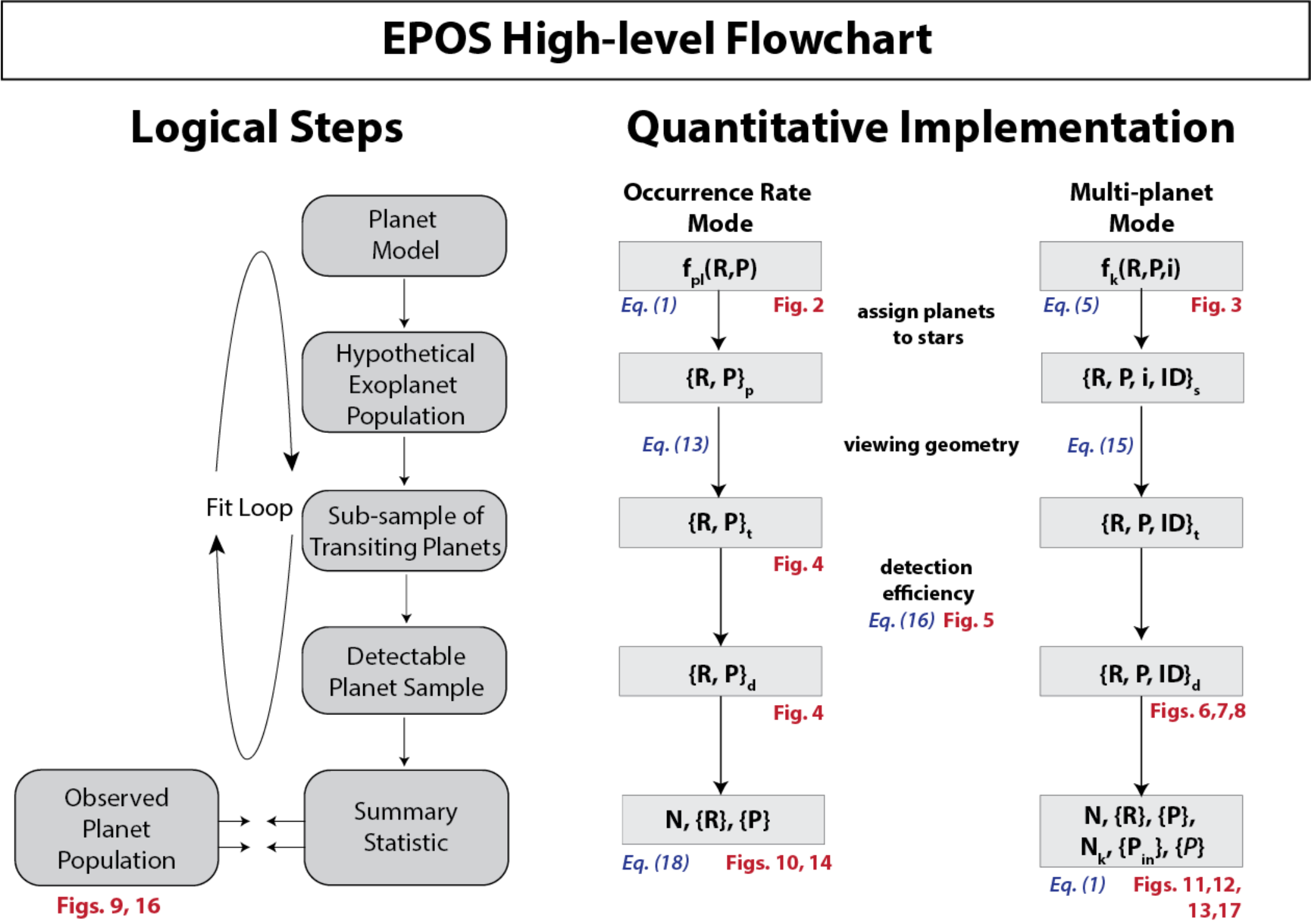}
    \caption{
    Flowchart of the Exoplanet Population Observation Simulator.
    A description of all mathematical symbols can be found in table \ref{t:sym}.
    }
    \label{f:flowchart}
\end{figure*}

In this paper, we verify this approach using parametric models of planet populations, that we compare to the final data release of the \kepler mission, \texttt{DR25}. 
From this, we are able to make a statistical evaluation of the properties of exoplanetary systems.
Among other results we report on the location of the innermost planet in planetary systems to place our Solar System in the context of exoplanet systems. We examine how the innermost planets in most of the \kepler systems are located much closer in than Mercury and Venus.
In an upcoming paper, we will use \epos to make a direct comparison between planet formation models and exoplanet populations.

\section{Code description}\label{s:code}
The Exoplanet Population Observation Simulator, \epos, employs a forward modeling approach to constrain exoplanet populations through the following iterative procedure:
\begin{description}
\item [Step 1] Define a distribution of planetary systems from analytic forms for planet occurrence rates and planetary architectures or from a planet population synthesis model.
\item [Step 2] From this derive a transiting planet population by assigning random orientations to each system and evaluating which planets transit their host stars.
\item [Step 3] Determine which of the transiting planets would be detected by \kepler, accounting for detection efficiency.
\item [Step 4] Compare the detectable planet population with exoplanet survey data using a summary statistic.
\item [Step 5] Repeat steps 1-4 until the simulated detectable planet population matches the observed planet population to constrain the intrinsic distribution of planetary systems.
\end{description}

In this section we will describe each step in greater detail.
Steps 1-4 take less than a second to run on a single-core CPU, allowing for an efficient sampling of the parameter space using MCMC methods.
Figure \ref{f:flowchart} summarizes these steps and their quantitative implementation in a flowchart. A description of all mathematical symbols used in this paper can be found in the Appendix, Table \ref{t:sym}.

\subsection{Step 1: Planet Distributions}\label{s:step:1}
Planet populations are generated using a Monte Carlo simulation by random draws from a multi-dimensional probability distribution function, $f$, which represents the intrinsic distribution of planets and planetary systems. We simulate a planet survey equal in size to the \kepler survey, roughly $160,000$ stars.
The parametric descriptions of $f$ are based on studies of planet occurrence rate and planet multiplicity from \kepler. Distributions based on planet formation models will be described in an upcoming paper. 

Here we describe two parametric descriptions for the planet population $f$ that correspond to two different modes in \epos:
\begin{description}
\item[Occurrence rate mode] Simulates only the distribution of planet radius, $R$, and orbital period, $P$, as $f=\fpl(R,P)$. This mode is similar to occurrence rate calculations that estimate the average number of planets per star, $\eta$, in a certain period and radius range.
This mode can also be used to estimate the planet mass distribution for comparison with microlensing data, see \cite{2018arXiv180300777P} for details. 
\item[Multi-planet mode] Simulates multiple planets per system, taking into account their relative spacing and mutual inclination, $\Delta i$, to estimate orbital architectures. The properties of each planet in the system are drawn from a distribution $f=f_k(R,P,i)$, where $i$ is the planet's orbital inclination and $k$ is an index for each planet in the system.
\end{description}

\begin{figure} % Fig 2
	%cp ~/EOStool/png/init_single/input/panels.png png/
      \includegraphics[width=\linewidth]{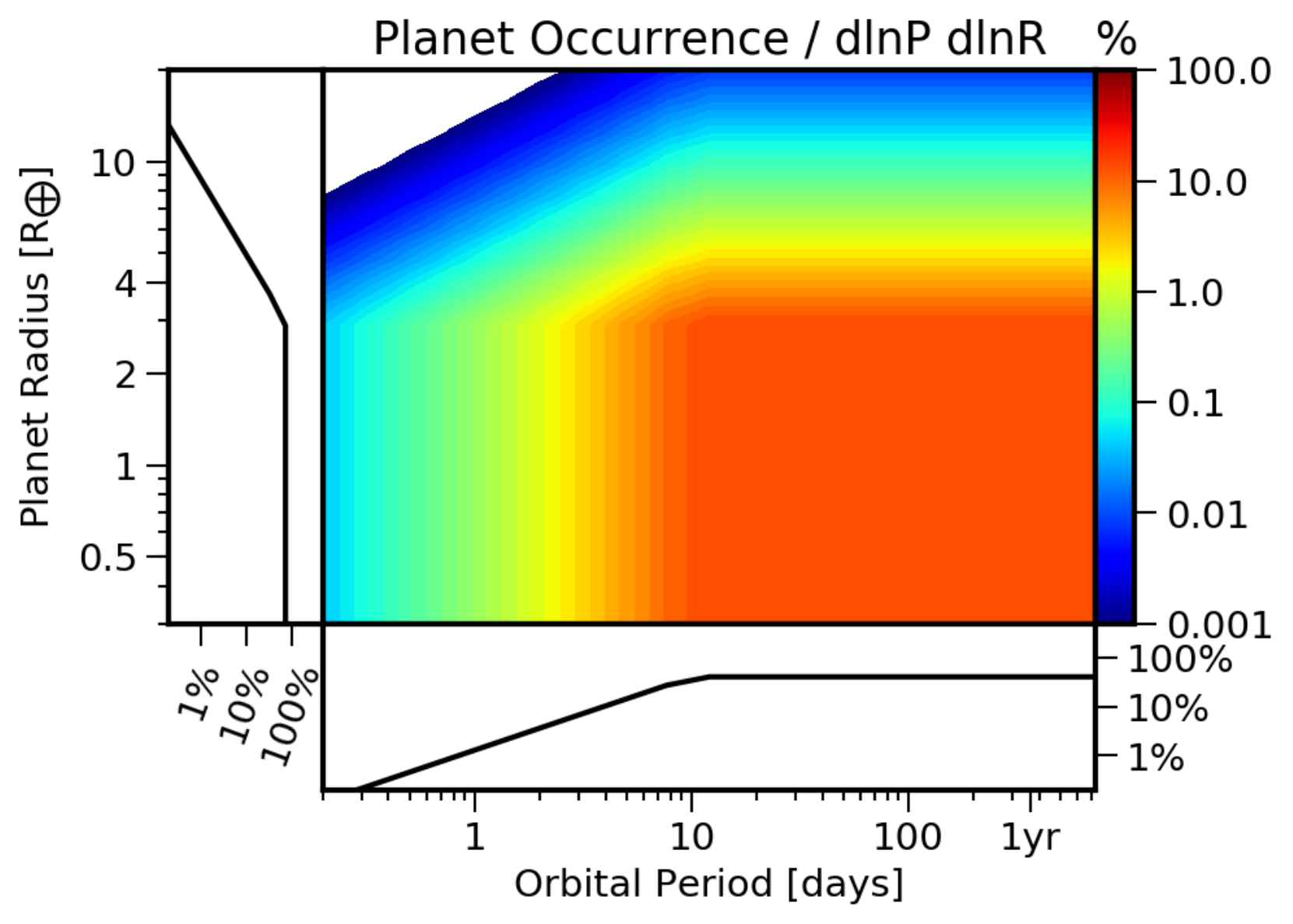}
        \caption{Example planet probability distribution function, $\fpl(P,R)$. 
    The color indicates the planet occurrence rate in percent per unit area of $d\ln P d\ln R$.
    The side panels show the marginalized distributions, $f_P(P)$ in units of $d\ln P$ and $f_R(R)$ in units of $d\ln R$. 
    %Break at $P=11$ days, $R=3.0 ~R_\oplus$, and $\eta=2.3$. [add marginalized P, R distributions?]
    }
    \label{f:panels}
\end{figure}

\subsubsection{Occurrence rate mode}\label{s:para}
The central assumption we use in this paper is that the planet occurrence rate distribution is a separable function in planet radius and orbital period,
\begin{equation}\label{eq:pdf}
\fpl(P,R) \propto f_P(P) f_R(R).
\end{equation}
The distribution is normalized such that integral over the simulated period and planet radius range equals the number of planets per star, $\eta$:
\begin{equation}
\eta=\int_P \int_R \fpl(P,R) ~d{\log}P ~d{\log}R 
\end{equation}

The planet orbital period distribution is described by a broken power law
\begin{equation}\label{eq:period}
    f_P(P)= 
\begin{cases}
    (P/P_{\rm break})^{a_P},& \text{if } P < P_{\rm break}\\
    (P/P_{\rm break})^{b_P},& \text{otherwise}
\end{cases}
\end{equation}
where the break reflects the observed flattening of sub-Neptune occurrence rates around an orbital period of ten days \citep[e.g.][]{2012ApJS..201...15H,2015ApJ...798..112M}. In this example we use $a_P=1.5$ for the power law index of the increase interior to $P_{\rm break}=10$ days, and a flat occurrence rate $b_P=0$, at longer periods but we will refine these values later.

The planet radius distribution function is similarly described by a broken power law
\begin{equation}\label{eq:radius}
    f_R(R)= 
\begin{cases}
    (R/R_{\rm break})^{a_R},& \text{if } R < R_{\rm break}\\
    (R/R_{\rm break})^{b_R},& \text{otherwise}
\end{cases}
\end{equation}
reflecting the observed increase of small planets with decreasing radius ($b_R=-4$), and the ``plateau'' ($a_R=0$) of constant occurrence rates for planets smaller than $R_{\rm break}=3 ~R_\oplus$ \citep[e.g.][]{2012ApJS..201...15H, 2013ApJ...770...69P}. 

The probability distribution function of planet occurrence thus has 7 free parameters ($\eta$, $P_{\rm break}$, $a_P$, $b_P$, $R_{\rm break}$, $a_R$, $b_R$). An example for $\eta=2$ is shown in Figure \ref{f:panels}.

\subsubsection{Multi-planet mode}\label{s:multi}
In its multi-planet mode \epos simulates a planetary system for each star. Each planet in the system, denoted by subscript $k$, is characterized by its radius, orbital period, and orbital inclination with respect to the observer, $i$, and there are $m$ planets per system.
Because the properties of planets in multi-planet systems tend to be correlated, we draw a typical set of planet properties for each system ($R_s,P_s,i_s$) on which the properties of each planet in the system ($R_k,P_k,i_k$) are dependent:
\begin{equation}\label{eq:pdf:multi}
\begin{aligned}
f_{\text{pl},k}(R,P,i|R_s,P_s,i_s) \propto
&f_{R,k}(R|R_s)\\
&f_{P,k}(P|P_s)\\
&f_{i,k}(i|i_s).
\end{aligned}
\end{equation}
where we make the assumption that distributions of planet size, period, and inclination are not interdependent.
Previous studies have shown that planets in multi-planet systems tend to be more similar in size and more regularly spaced than random pairings from the overall distribution \citep{2017ApJ...849L..33M,2018AJ....155...48W} and have low mutual inclinations of a few degrees \citep[e.g.][]{2014ApJ...790..146F}. This motivates our choice to not draw the properties of each planet independently from a distribution $\fpl(R,P,i)$. 

\begin{figure} % Fig 3
\includegraphics[width=\linewidth]{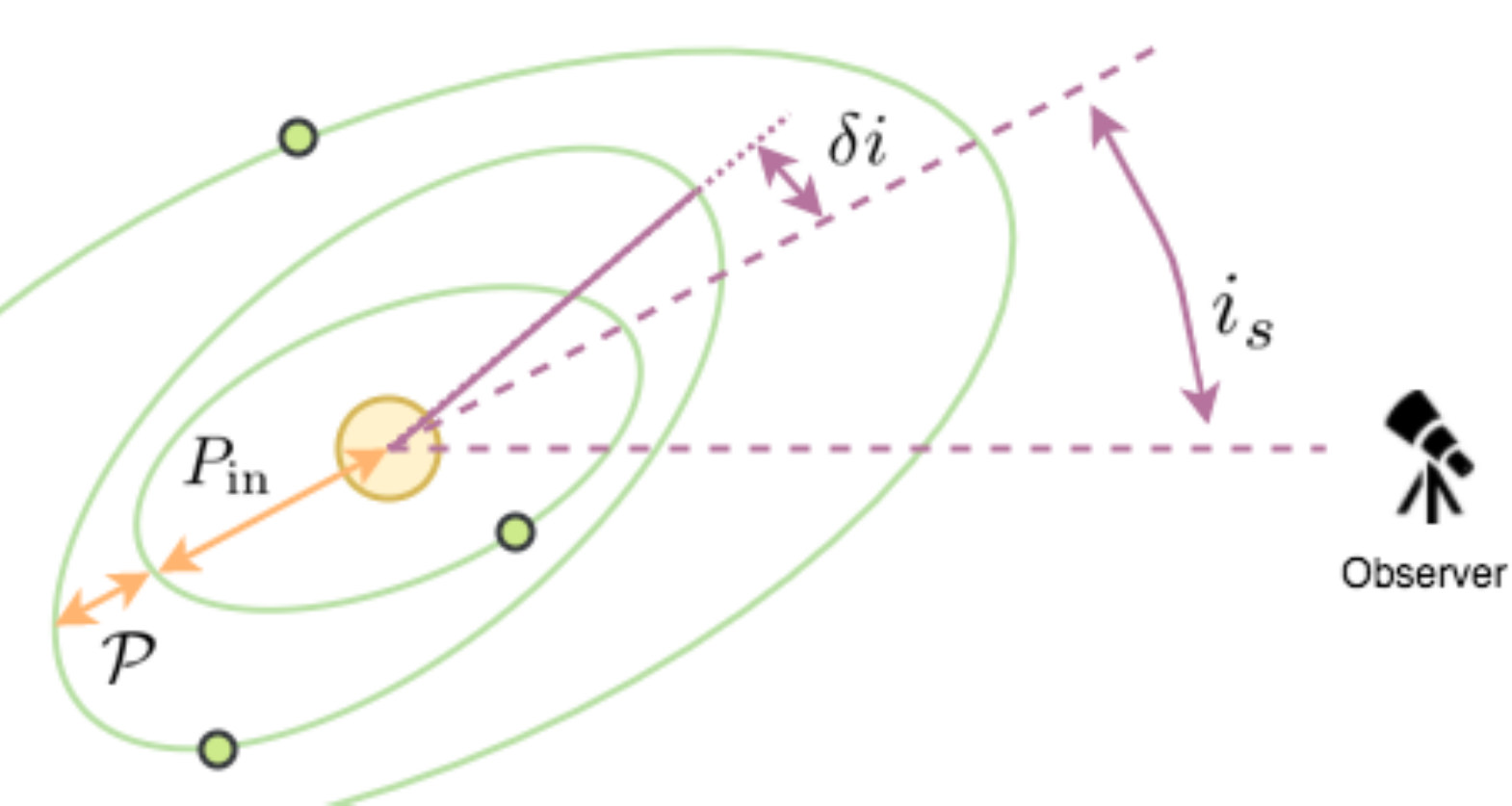}
\caption{Illustration of the planetary system architecture. 
$P_\text{in}$ denotes the orbital period of the innermost planet, while $\mathcal{P}$ denotes the period ratio between adjacent planets.
The system is inclined with respect to the observer by an angle $i_s$,
and each planet has a mutual inclination $\delta i$ with respect to this system inclination. 
Note that each planet has a different mutual inclination and therefore a different inclination with respect to the line of sight. 
All orbits are assumed to be circular and the ellipiticity of the projected orbits are due to their respective inclinations.
The probability distributions for $i_s$, $\delta i$, $\mathcal{P}$, and $P_\text{in}$ are described in the text.
\label{f:art}
}
\end{figure}

The system properties ($R_s,P_s,i_s$) are drawn from a distribution
\begin{equation}
g_\text{pl}(R_s,P_s,i_s) \propto
g_R(R_s) g_P(P_s) g_i(i_s)
\end{equation}
where once again we made the assumption that the distributions of planet size, period, and inclination are separable functions of each variable. The distribution of $g$ is normalized such that the integral over the simulated parameter range is equal to the fraction of stars with planetary systems, $\eta_s$:
\begin{equation}
\eta_s=\int_P \int_R \int_i g_\text{pl}(P,R,i) ~d{\log}P ~d{\log}R ~di
\end{equation}
The parameterization of the system and planet properties are described below. The system architecture is illustrated in Figure \ref{f:art}

\paragraph{Orbital Inclination}
A planet's orbital inclination with the respect to the line of sight, $i_p$, is dependent on the inclination of the system, $i_s$, the mutual inclination of the planet with respect to that system, $\delta i$, and the longitude of the ascending node, $\delta \Omega$.

We assume that the planet mutual inclinations, $\delta i$, follow a Rayleigh distribution: 
\begin{equation}\label{eq:pldeltainc}
f_{\delta i,k}(\delta i)= \delta i/\Delta i e^{-\delta i^2/(2 {\Delta i}^2)},
\end{equation}
where $\Delta i$ is the mode of the mutual inclination distribution of planetary orbits, which is typically $1-3\degr$ \citep{2014ApJ...790..146F}. The distribution of system inclinations with respect to the observer, $i_s$, is proportionate to $\cos(i)$. 
\begin{equation}\label{eq:sysinc}
g_i(i_s) \propto \cos(i_s)
\end{equation}
where $i_s=0$ is an edge-on orbit.
The distribution of planet inclinations with respect to the observer is then
 \begin{equation}\label{eq:plinc}
f_{i,k}(i|i_s)= |i_s + f_{\delta i,k}(\delta i) \cos(\pi \delta \Omega)| 
\end{equation}
where $\delta \Omega$ is the longitude of the ascending node with respect to the observer.

\paragraph{Orbital Period}
We assume that the planets are regularly spaced in the logarithm of the orbital period, 
with the location of the first planet ($k=0$) defining the location of additional planets ($k=1...m$) in the system. The period ratio of adjacent planets is denoted by $\mathcal{P}_k \equiv P_{k+1}/P_{k}$ where $P_{k+1}$ is always the planet with the larger orbital period.

The location of the first planet in the system, ($P_0 \equiv P_s$), is parameterized by the broken power--law distribution described by Equation \ref{eq:period}, but with a steeper decline in planet occurrence with orbital period ($b_P\lesssim 0$) that we will constrain in the fitting process. 

Observed orbital spacings follow a broad range in period ratios, and we follow \cite{2015ApJ...808...71M} in parameterizing the distribution of dimensionless spacings, $D_k=2\frac{\mathcal{P}_k^{2/3}-1}{\mathcal{P}_k^{2/3}+1}$, as a log-normal distribution. 
The orbital period distribution of the k-th planet in the system is then given by
\begin{equation}\label{eq:dP}
f_{P,k}(P|P_{k-1})= \frac{1}{\sqrt{2 \pi} \sigma} e^{\frac{(\log D_k-D)^2}{2 \sigma^2}}
\end{equation}
with respect to the orbital period of the previous planet, $P_{k-1}$. $D$ and $\sigma$ are free parameters that characterize the median and width of the distribution, with typical values of $D\approx-0.4$ and $\sigma \approx 0.2$. 

\paragraph{Planet Radius}
We assume all planets in the system are of equal size
\begin{equation}
f_{R,k}(R|R_s)= g_R(R_s)
\end{equation}
where we assume the radius distribution follows Equation \ref{eq:radius}.
We choose equal-sized planets over randomly assigned sizes because planets in multi-planet systems tend to be of similar size \citep{2011ApJS..197....8L,2017ApJ...849L..33M,2018AJ....155...48W}. 
While there are observed trends of increasing planet size with orbital period, these trends are strongest at short orbital periods \cite[e.g.][]{2018arXiv180405069C}
and for large planet sizes \citep{2013ApJ...763...41C,2016ApJ...825...98H} that we will exclude from our observational comparison, see \S \ref{s:obs}.

Thus the number of free parameters describing the population of planetary systems is 11 ($\eta$, $m$, $\Delta i$, $P_\text{break}$, $a_P$, $b_P$, $D$, $\sigma$, $R_\text{break}$, $a_R$, $b_R$).

\begin{figure} % Fig 4
    \includegraphics[width=\linewidth]{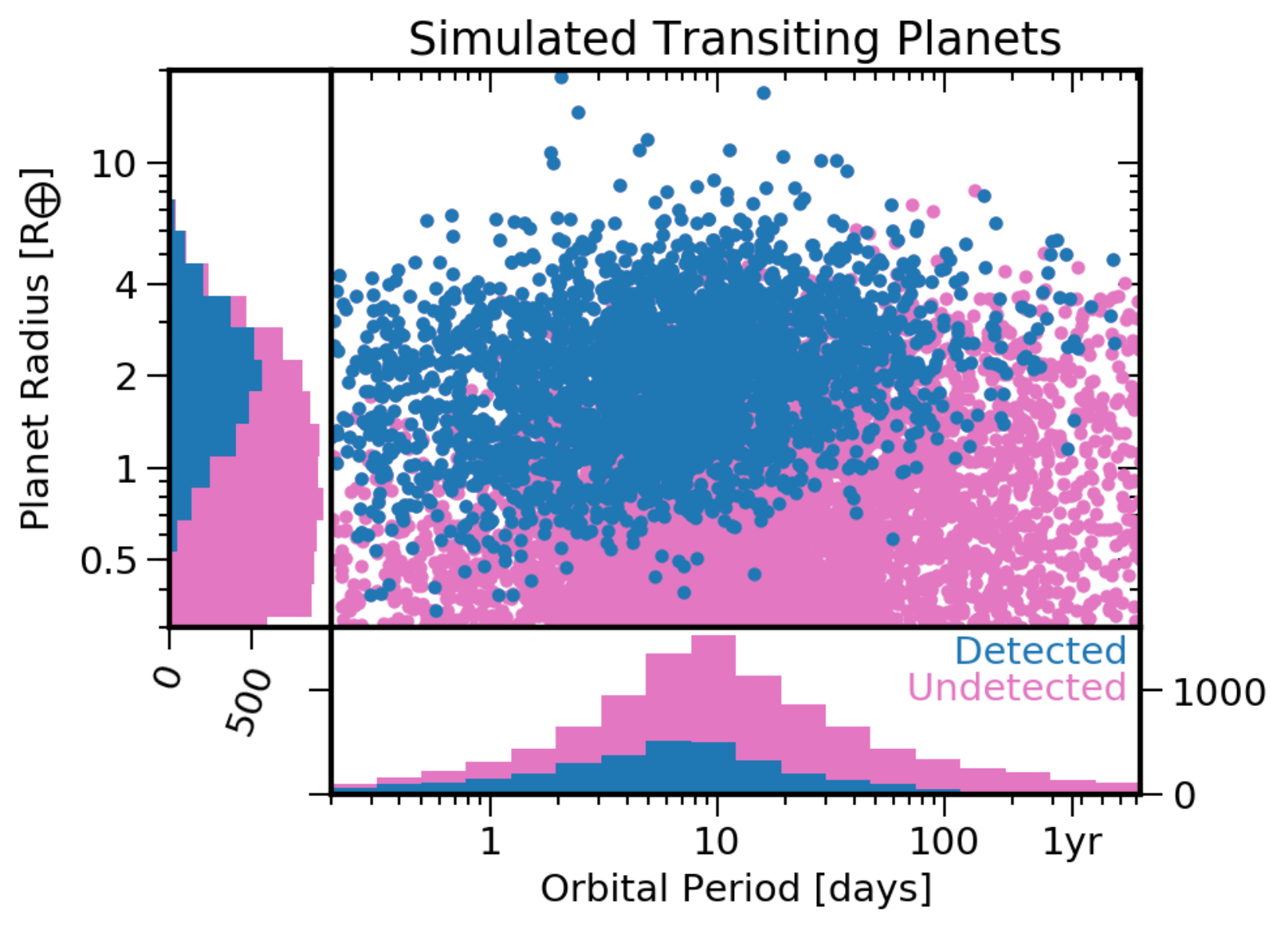}
    %cp ~/EOStool/png/init_single/output/periodradius.transit.png png/PR-transit.png
    \caption{Simulated sample of transiting (pink) and detectable (blue) planets generated in occurrence rate mode, see Figure \ref{f:panels}.
    }
    \label{f:PR:transit}
\end{figure}

\subsection{Step 2: Transiting Planet Populations}\label{s:step2} 
The synthetic planet population is simulated using a Monte Carlo approach by sampling the distribution function of planet properties (Eq. \ref{eq:pdf} or \ref{eq:pdf:multi}) $n_p$ times. We simulate planetary systems in the period range $[0.5,730]$ days and $[0.3,20] R_\oplus$ (Figure \ref{f:PR:transit}). For each planet, 2 random numbers are drawn to determine its properties. The first random number is used to determine the planet orbital period from the cumulative distribution function of Eq. \ref{eq:period}. The second random number is used to determine the planet radius from the cumulative distribution function of Eq. \ref{eq:radius}.

The total number of draws in the synthetic survey is $n_p=\eta n_\star$ where $n_\star$ is the number of stars in the survey ($n_\star \approx160,000$), multiplied by the average number of planets per star, $\eta$. In occurrence rate mode, with $\eta\approx2$, the simulated sample $\{R,P\}_p$ consists of $n_s\approx240,000$ planets. In multi-planet mode, with $\eta_s\approx0.35$ and $m\approx 6$, the simulated sample $\{R,P,i\}_s$
consists of the same number of planets distributed across $n_s\approx40,000$ systems. Each star in the survey is assigned a unique identifier, \ID, to keep track of the observable planet multiplicity.

In occurrence rate mode, the subset of transiting planets can be simulated by considering the geometric transit probability
\begin{equation}\label{eq:fgeo}
\fgeo = R_\star/a.
\end{equation}
The semi-major axis $a$ is calculated using Kepler's third law. The stellar mass and radius are the average values of the surveyed stars (see next section). 
A planet is transiting if
\begin{equation}\label{eq:mc:geo}
\chi < \fgeo  
\end{equation}
where $\chi$ is a continuous random variable between $0$ and $1$. The transiting planet sample, $\{R,P\}_t$, typically contains $10,000$ planets. An example is shown in Figure \ref{f:PR:transit}.

\subsubsection{Multi-planet transit probability}
The Monte Carlo simulation of transiting planetary systems goes as follows. 
First, the system inclination, $i_s$ is drawn according to Equation \ref{eq:sysinc}. The draw from a distribution is performed by generating a random number between 0 and 1 and interpolate the parameter value from the normalized cumulative distribution. 
Next, the mutual inclination of the orbit of each planet in the system, $\delta i_{p,k}$, is drawn from Equation \ref{eq:pldeltainc}. Then the longitude of the ascending node with respect to the observer, $\delta \Omega$, is randomly drawn for each planet from a uniform distribution, $\delta \Omega= \chi$.  The inclination, $i_k$ of each planets orbit with respect to the line-of-sight can then be calculated from equation \ref{eq:plinc}.
The transiting planet population, $\{R,P\}_t$, is defined by planets that traverse the stellar limb, given by
\begin{equation}\label{eq:multi:fgeo}
i_k<\arcsin(\fgeo),
\end{equation}
where \fgeo is the geometric transit probability from Eq. \ref{eq:fgeo}. We note that for a single planet ($\delta i_{p,k}=0$), this expression is equivalent to the geometric transit probability of Eq. \ref{eq:fgeo}. This expression is only valid for small mutual inclinations.

To account for the \textit{Kepler dichotomy}, the apparent excess of single transiting systems \citep[e.g.][]{2016ApJ...816...66B}, we assume that a fraction \fiso of planetary systems has an isotropic distribution of orbits described by Eq. \ref{eq:fgeo} instead of Eq. \ref{eq:multi:fgeo}. Typically, $\fiso \approx 0.5$, but we will treat it as a free parameter to be constrained from the data.

\begin{figure*} %Fig 5
	% cp ~/EOStool/png/init_single/survey/efficiency.novet.png png/efficiency.png
	% cp ~/EOStool/png/init_single/survey/vetting.png png/
	% cp ~/EOStool/png/init_single/survey/completeness.png png/
	\includegraphics[width=0.33\linewidth]{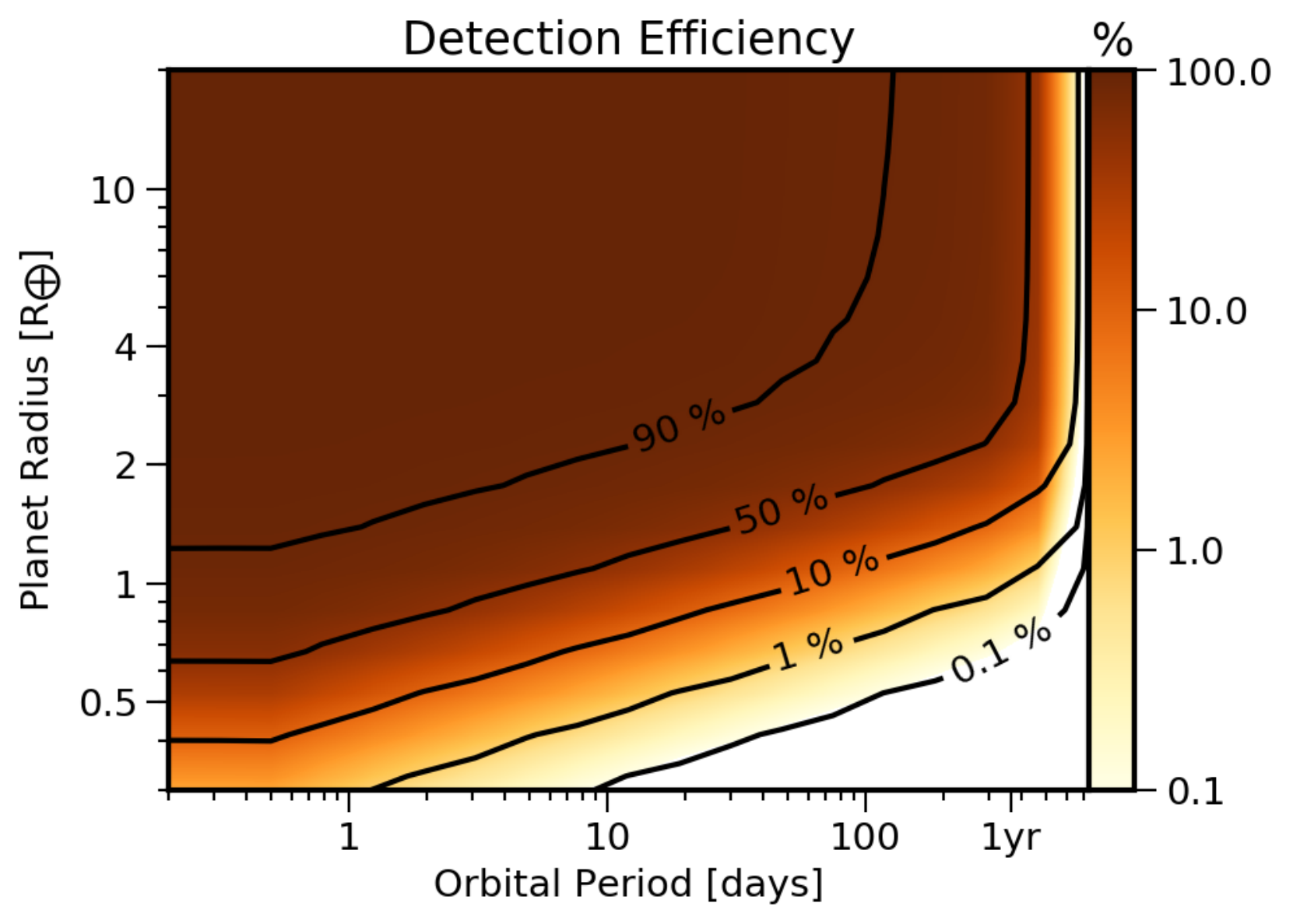}
	\includegraphics[width=0.33\linewidth]{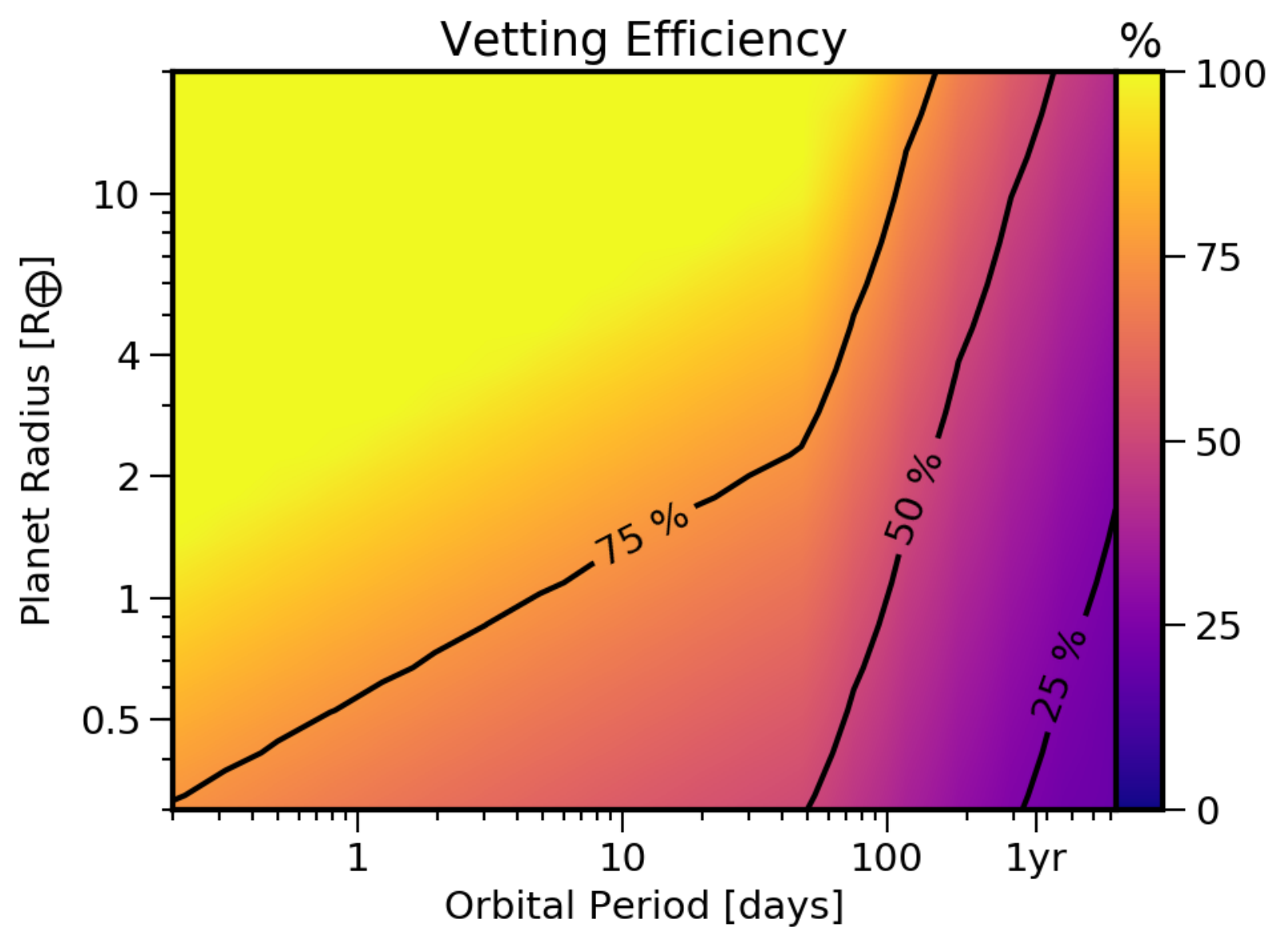}
    \includegraphics[width=0.33\linewidth]{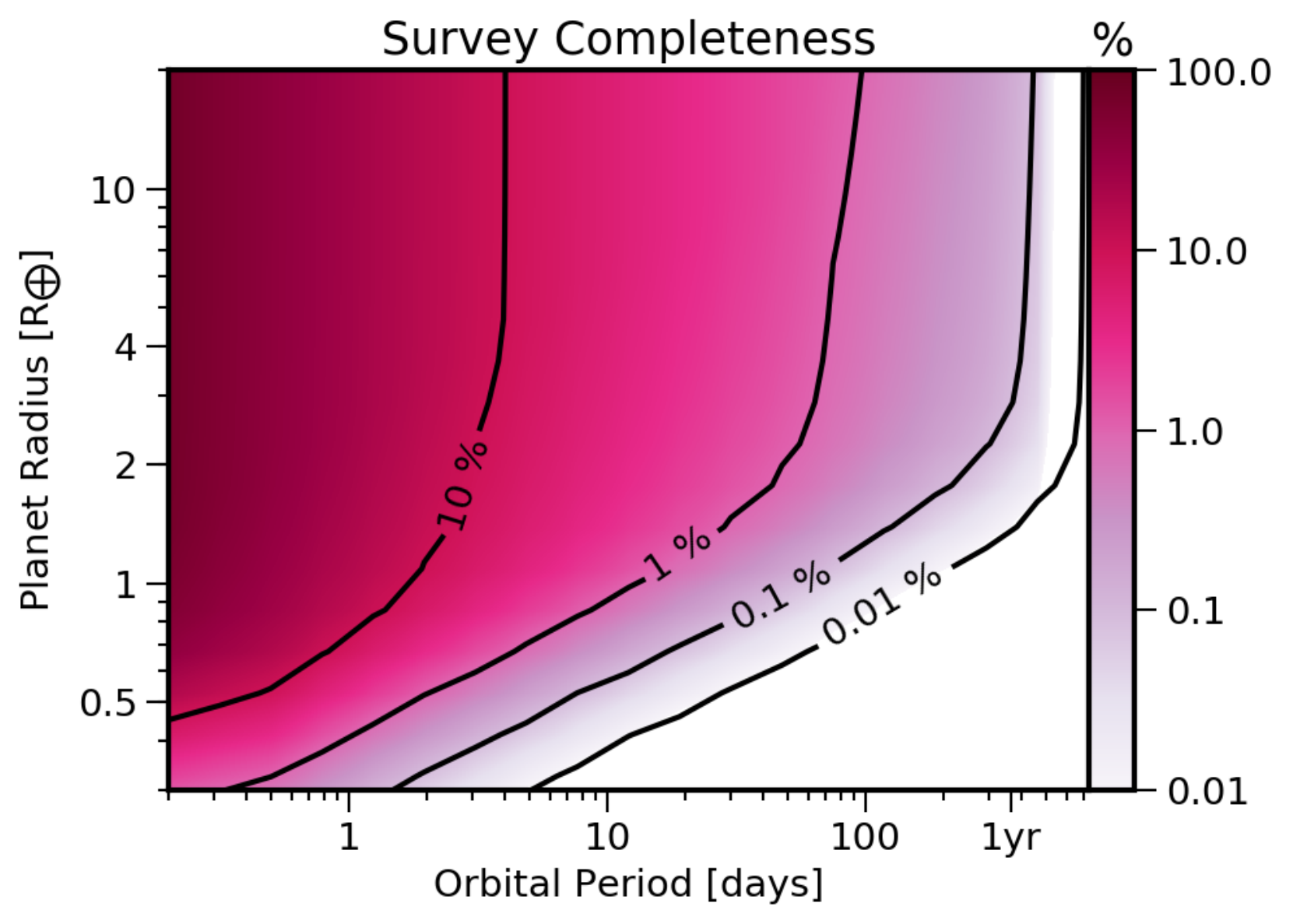}
    \caption{Planet detection efficiency for dwarf stars from the \kepler mission for the \texttt{DR25} data release as function of orbital period and planet radius.
    Left Panel: The detection efficiency, \fsnr, the probability that a planet candidate is detected based on the signal-to-noise of the transit. 
    Middle Panel: The vetting efficiency, \fvet, the probability that a planet is classified as a reliable planet candidate (a \textit{Robovetter} disposition score of $\geq 0.9$).
    Right Panel: The survey completeness, \fdet, which includes both the detection and vetting efficiency as well as the the geometric transit probability, $\fdet= \fgeo ~\fsnr ~\fvet$}.
    \label{f:eff}
\end{figure*}

\subsection{Step 3: Survey Detection Efficiency}\label{s:step3}
We simulate a detectable planet sample, $\{R,P\}_d$, from the simulated sample of transiting planets,  $\{R,P\}_t$, by taking into account the survey detection efficiency, $\fsnr(R,P)$, and the vetting completeness, $\fvet(R,P)$, described below and displayed in Figure \ref{f:eff}. A planet is detectable if:
\begin{equation}\label{eq:snr}
\chi < \fsnr \fvet
\end{equation}
where $\chi$ is a continuous random variable between 0 and 1. $\fsnr(R,P)$ is the detection efficiency based on the combined signal-to-noise ratio of all planet transits. $\fvet(R,P)$ is the detection efficiency of the \kepler \textit{Robovetter} for a planet candidate sample with high reliability.

We calculate detection efficiency contours for each individual star observed by \kepler using \texttt{KeplerPORTs}\footnote{\url{https://github.com/nasa/KeplerPORTs}} \citep{2017ksci.rept...19B}. 
We included all stars that were fully searched by the Kepler pipeline, for which stellar properties were available in the \cite{2017ApJS..229...30M} catalog, and for which all detection metrics were available on the exoplanet archive\footnote{\url{https://exoplanetarchive.ipac.caltech.edu/docs/Kepler_completeness_reliability.html}}. 

The detection efficiencies were evaluated on a grid with 20 logarithmically-spaced orbital period bins between $0.2$ and $730$ days assuming planets on circular orbits, and 21 logarithmically spaced planet radius bins between $0.2$ and $20 ~R_\oplus$.
After removing giant and sub-giant stars according to the effective temperature dependent surface gravity criterion in \cite{2016ApJS..224....2H}, the sample consists of $n_\star= 159,238$ stars, with a median mass of $M_\star=0.95 ~M_\odot$ and median radius of $R_\star=0.94 ~R_\odot$. 
We calculate the survey detection efficiency, $\fsnr(R,P)$, by averaging the individual contributions of each star, displayed in the upper panel of Figure \ref{f:eff}.

Not all transiting planets that are detectable based on the detection efficiency are vetted as planet candidates by the \kepler \textit{Robovetter} \citep{2016ApJS..224...12C}. Including only reliable planet candidates, here defined with a disposition score larger than $0.9$, further reduces the vetting completeness. 
The vetting efficiencies were calculated following \cite{2018ApJS..235...38T} based on the \kepler simulated data products\footnote{\url{https://exoplanetarchive.ipac.caltech.edu/docs/KeplerSimulated.html}}.
The vetting efficiency was evaluated on the same radius and period grid as the detection efficiency and using the same criterion to select main-sequence stars. 
We use a power-law in planet radius and a broken power-law in orbital period to obtain a smooth function for the vetting completeness \fvet
\begin{equation}
\fvet= c ~R^a_R
\begin{cases}
    (P/P_{\rm break})^{a_P},& \text{if } P < P_{\rm break}\\
    (P/P_{\rm break})^{b_P},& \text{otherwise}
\end{cases}
\end{equation}
with $c=0.63$, $a_R= 0.19$, $P_{\rm break}=53$, $a_P=-0.07$, $b_P=-0.39$ (see Appendix \ref{s:vet} for details).
The best-fit vetting efficiency is shown in Figure \ref{f:eff}. The vetting efficiency varies between close to 100\% for Neptune-sized planets at short orbital periods to 25\% near the habitable zone.
The break in the power-law is needed to match the reduced vetting efficiency for planets at orbital periods larger than $\sim 100$ days as described in \cite{2018ApJS..235...38T}.

\begin{figure} % Fig 6
	% cp ~/EOStool/png/init_multi/output/PR.multi.png png/f_PR_multi.png
    \includegraphics[width=\linewidth]{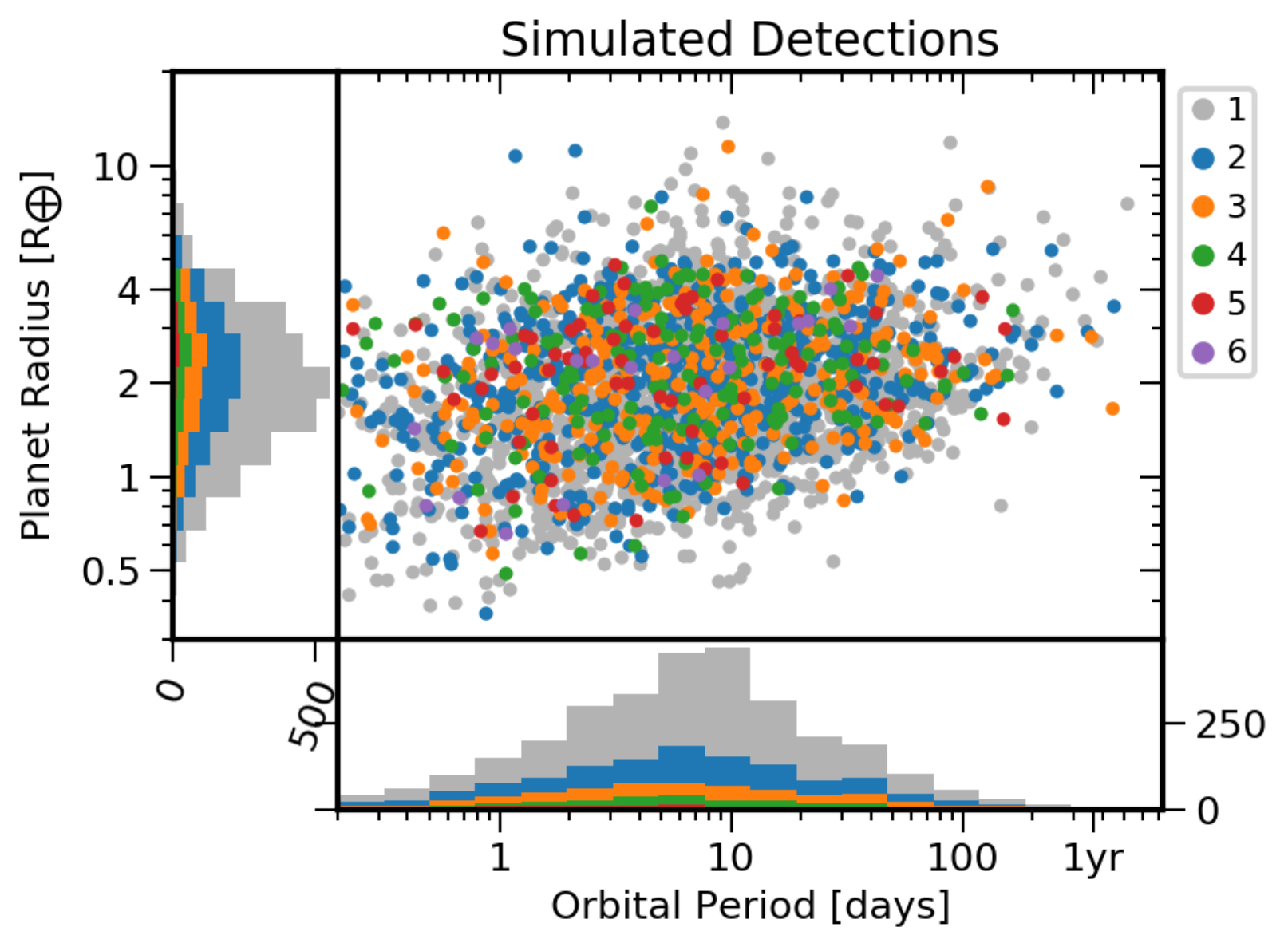}
    \caption{Simulated sample of detectable planetary systems. Planets with no additional planets detected in the system are color-coded in gray.
    Colors indicate the number of observed planets per system.
    }
    \label{f:PR:multi}
\end{figure}

\begin{figure} % Fig 7
	% cp ~/EOStool/png/init_multi/output/periodratio.index.input.png png/f_in_dP.png
    \includegraphics[width=\linewidth]{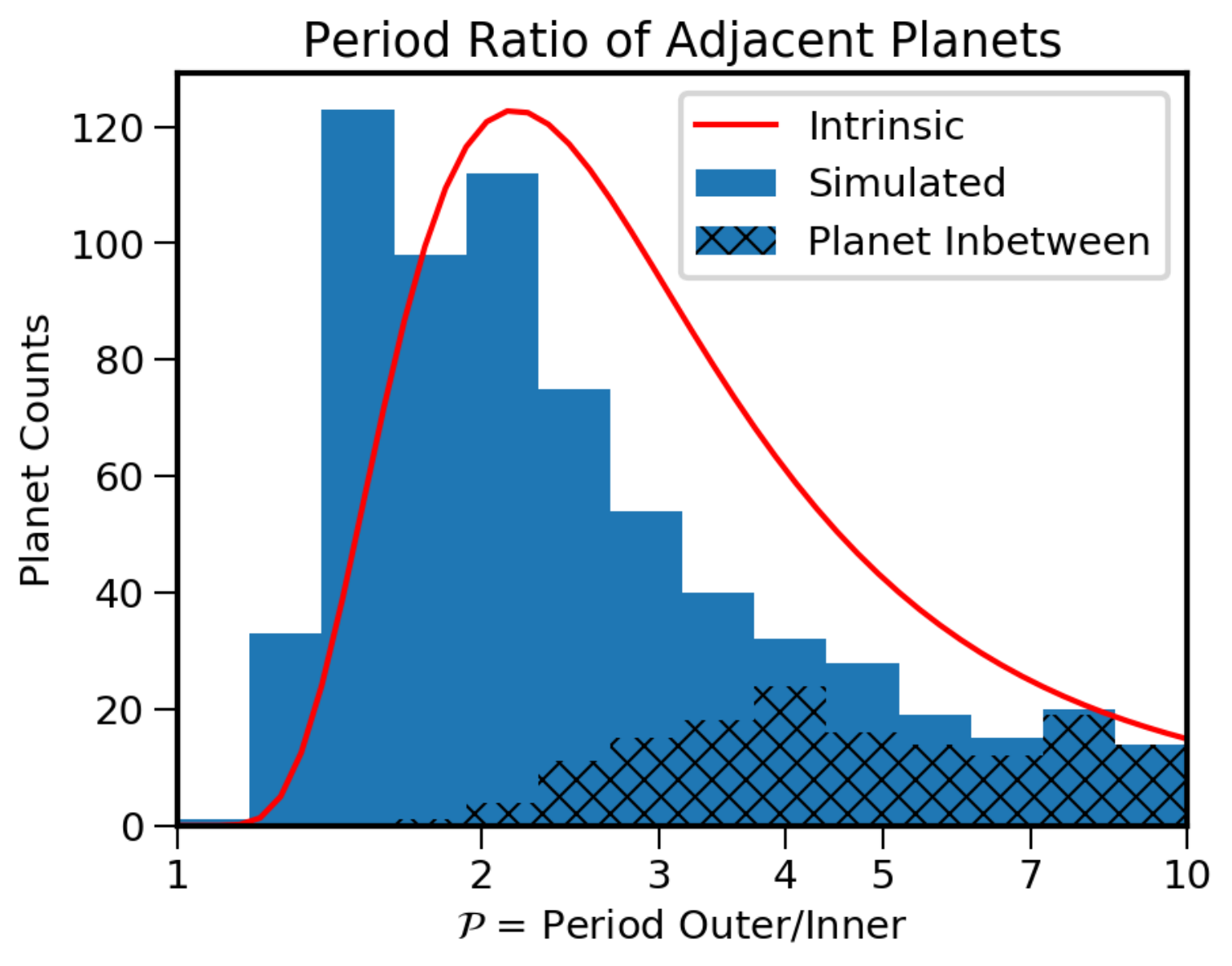}
    \caption{Period ratio distribution of adjacent planets in multi-planet systems, $\mathcal{P}=P_{k+1}/P_k$. The intrinsic distribution is shown with the red line while the solid blue histograms show the distribution of detections in the simulated survey. The observable distribution is skewed towards shorter orbital period ratios by detection biases. 
    The hatched region indicates the observable planet pairs with at least one non-detected planet between them.
    }
    \label{f:in:dP}
\end{figure}

\begin{figure} % Fig 8
	% cp ~/EOStool/png/init_multi/output/innerperiod.input.png png/f_in_Pinner.png
     \includegraphics[width=\linewidth]{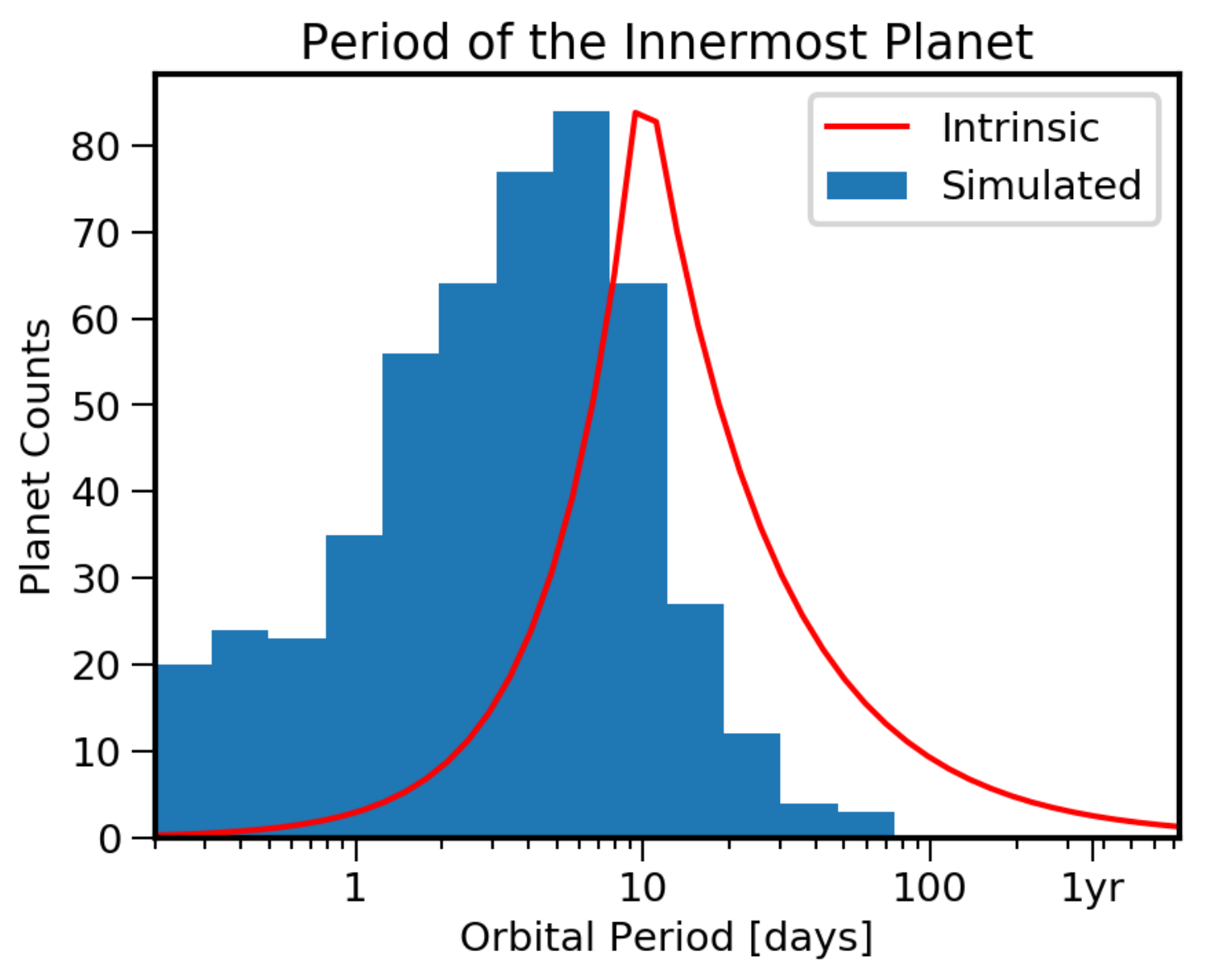}
    \caption{Orbital period distribution of the innermost planet in each system. The intrinsic distribution is shown with the red line while the solid blue histograms show the distribution of detections in the simulated survey. The observable distribution is skewed towards shorter orbital periods by detection biases. 
    }
    \label{f:in:Pinner}
\end{figure}

The detectable planet sample, $\{R,P\}_d$, typically contains about $4,000$ planets (Figure \ref{f:PR:transit}). In multi-planet mode, \epos also keeps track of which planets are observable as part of multi-planet systems (Figure \ref{f:PR:multi}). From the observed multi-planet systems we generate a set of summary statistics: $N_k$, the number of stars with $k$ detectable planets;
$\mathcal{P}$, the orbital period ratio between adjacent planets (Fig. \ref{f:in:dP}); $P_\text{in}$, the orbital period of the innermost planet in the system (Fig. \ref{f:in:Pinner}).

It is worth noting that the properties of the observable multi-planet systems are significantly different from the distributions from which they are generated. When the intrinsic population contains only planetary systems with $m \geq 7$ planets, the majority of systems have $k=2-6$ transiting detectable planets. In addition, the observable period and period ratio distributions are skewed by detection biases. 

The geometric transit probability favors the detection of planets at shorter orbital periods. The distribution of the location of the innermost {\em observed} planet in the systems peaks at $P_\text{in}\sim 6$ days compared to $P_\text{in}= 10$ days in the intrinsic distribution (Fig. \ref{f:in:Pinner}). We also find that planetary systems at very short orbital periods ($P_\text{in}\sim 1$ day) are overrepresented in the detectable distribution by an order of magnitude, while systems with orbital periods similar to the terrestrial planets ($P_\text{in}\sim 100$ days) are underrepresented by an order of magnitude, due to the same detection biases (Fig. \ref{f:in:Pinner}).

Pairs of planets with smaller orbital period ratios are more likely to be both transiting, shifting the peak of the observable orbital period ratio distribution to smaller period ratios (Fig. \ref{f:in:dP}). The hatched area in the bottom panel of Figure \ref{f:in:dP} shows simulated planet pairs that are observable as adjacent but have a non-transiting planet between them. These planet pairs dominate the period ratio distribution at large orbital period ratios ($\mathcal{P}\gtrsim 4$).

\subsection{Step 4: Observational Comparison}\label{s:obs}
In this step we compare the simulated planet populations to the observed exoplanet properties to evaluate how well the simulated planet population reproduces the collection of observed planetary systems. We generate a set of summary statistics for both the simulated data and the \kepler survey. 
We evaluate the survey as a whole, and do not consider dependencies on stellar properties such as stellar mass \citep{2015ApJ...798..112M,2015ApJ...814..130M} or metallicity \citep{2016AJ....152..187M,2018AJ....155...89P}.

In occurrence rate mode, the summary statistics are the planet radius distribution, $\{R\}_d$, the orbital period distribution $\{P\}_d$, and the total number of planets, $N$. In multi-planet mode, additional summary statistics are calculated for the number of stars with $k$ planets, $N_k$, the period ratio distribution between adjacent planets, $\mathcal{P}$, and the period of the innermost planet in the system, $P_\text{in}$.
We evaluate this summary statistic in the range 
of $R=[0.5,6] R_\oplus$ and $P=[2,400]$ days. These ranges exclude two regions where the assumptions of separability of parameters clearly breaks down. The maximum planet size of $6 R_\oplus$ is chosen to exclude giant planets, which have a different distribution of orbital periods than sub-Neptunes \citep{2013ApJ...778...53D,2016A&A...587A..64S}, are less often part of multi-planet systems \citep{2012PNAS..109.7982S}, or have very dissimilar sizes than other planets in the system \citep{2016ApJ...825...98H}. The minimum orbital period of 2 days is chosen to exclude the photo-evaporation desert \citep[e.g.][]{2016NatCo...711201L} where the planet-radius distribution deviates significantly from that at larger orbital periods. The other bounds are chosen because there are very few planet detection outside this range. 

\begin{figure} % Fig 9
	% cp ~/EOStool/png/init_multi/survey/{PR.multi}.png png/f_multi_planets.png
    \includegraphics[width=\linewidth]{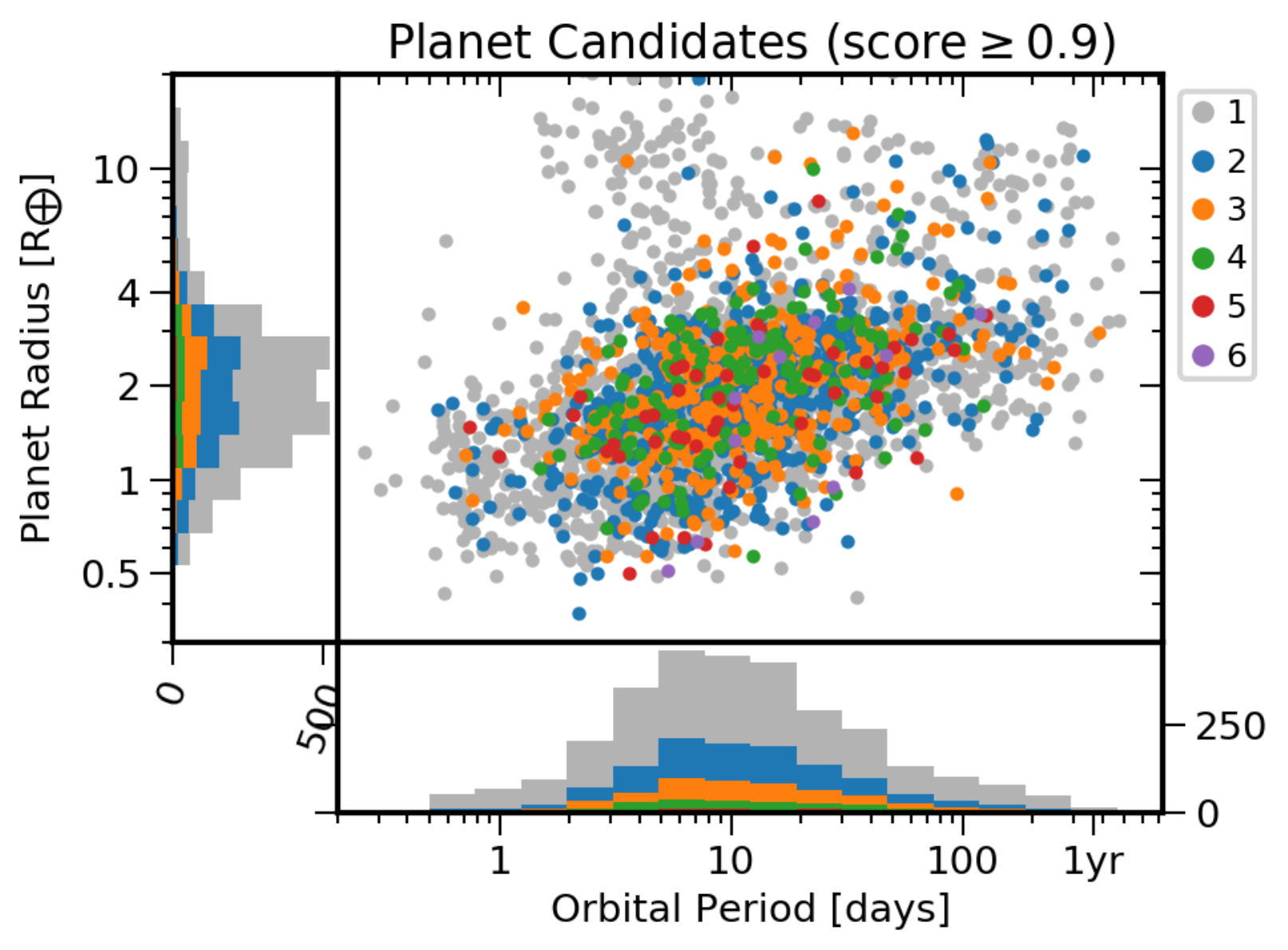}
    \caption{Observed sample of planetary systems. Planets with no additional planets detected in the system are color-coded in gray. Colors indicate the number of observed planets per system.
    }
    \label{f:multi:planets}
\end{figure}

We then generate a summary statistic from the \textit{Kepler} \texttt{DR25} catalog.
The planet candidate list is taken from \cite{2018ApJS..235...38T}.
We include only main sequence planet hosts by removing giant and sub-giant stars according to the effective temperature dependent surface gravity criterion in \citep{2016ApJS..224....2H}. We also use a disposition score cut of 0.9 to select a more reliable sample of planet candidates (see \citealt{2018ApJS..235...38T} for details). 
We account for the lower completeness of this high-reliability planet sample by explicitely taking into account the vetting completeness in the calculation of the survey detection efficiency.

The final list containing 3,041 planet candidates is shown in Figure \ref{f:multi:planets}. 
The list contains 1,840 observed single systems and 324 double, 113 triple, 38 quadruple, 10 quintuples, and 2 sextuple systems within the region where the summary statistic is evaluated.

\begin{figure} % Fig 10
	% cp ~/EOStool/png/zoom_single/output/pdf.zoom.png png/f_PR_single.png
    \includegraphics[width=\linewidth]{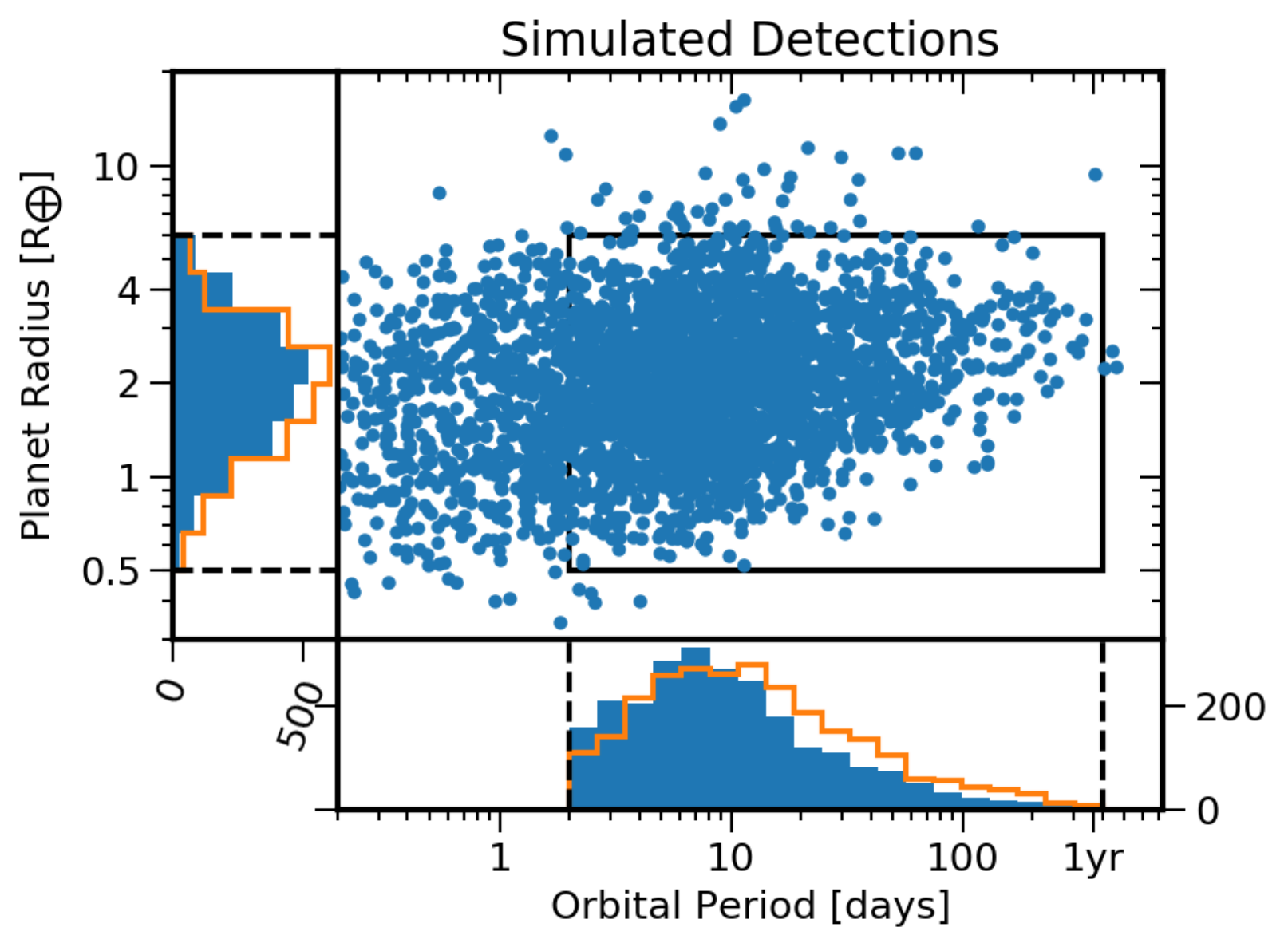}
    \caption{Comparison of simulated planets for the example model (blue) with detected planets (orange). The comparison region (black box) excludes hot Neptunes ($P\textless 2$ days) and giant planets ($R>6 R_\oplus$).
    }
    \label{f:PR:single}
\end{figure}

\subsubsection{Occurrence rate mode}\label{s:obs:occ}
Figure \ref{f:PR:single} shows how the summary statistics of planet radius and orbital period are generated from the detectable planet population (blue) and from the \kepler exoplanet population (orange). We compare the planet radius distributions and orbital period distributions separately. While this approach ignores any covariances between planet radius and orbital period that are present in the \kepler data, it is consistent with the assumption made in Eq. \ref{eq:pdf} that these functions are separable. We quantify the distance between the two distributions using the two-sample Kolmogorov-Smirnoff (KS) test and calculated the associated probabilities, $p_P$ and $p_R$, that the observed and simulated distributions are drawn from the same data.
We will minimize the differences between these distributions in the fitting step.

\begin{figure} % Fig 11
    % cp ~/EOStool/png/zoom_multi/output/multiplicity.planets.png png/f_multi_obs.png
    \includegraphics[width=\linewidth]{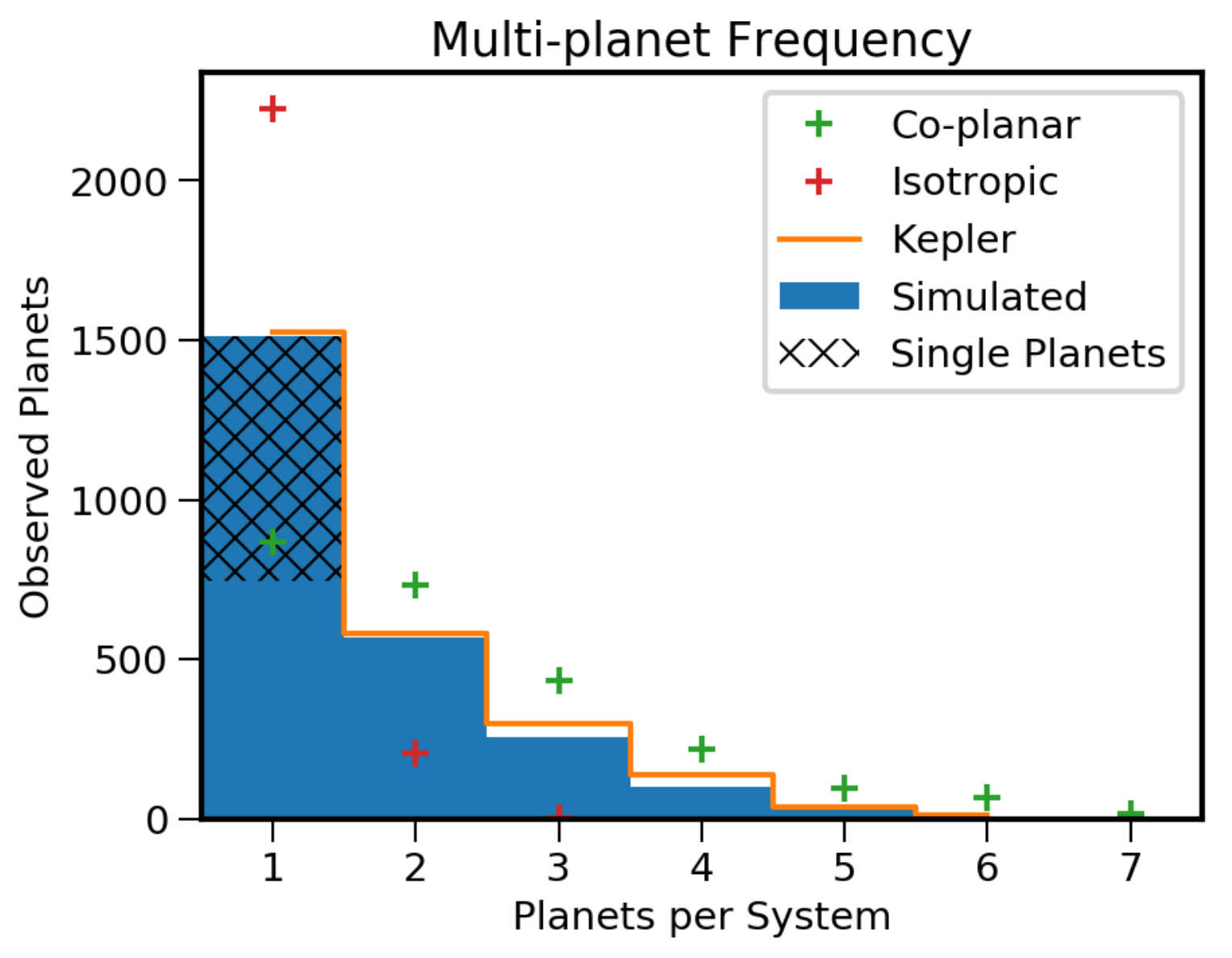}
    \caption{
    Simulated versus observed frequency of multi-planet systems. The blue histogram shows the example model with an average mutual inclination of $\Delta i=2$.
    Multi-planet statistics from Kepler derived in the same radius and period range are shown in orange. 
    The hatched region indicates the excess of single transiting planets, here $40\%$ of systems ($\fiso=0.4$).
	Crosses indicate a population of planetary systems on co-planar orbits (green, $\Delta i=0$) and on isotropic orbits (red, $\fiso= 1.0$).
    }
    \label{f:multi:obs}
\end{figure}

\subsubsection{Multi-planet statistics}\label{s:obs:multi}
\begin{figure} % Fig 12
    %\includegraphics[width=0.49\linewidth]{\string~/EOStool/png/zoom_multi/output/{periodratio}.png}
    % ./paper_zoom.py multi pair
    % cp ~/EOStool/png/zoom_multi/output/periodratio.png png/f_multi_ratio.png
    \includegraphics[width=\linewidth]{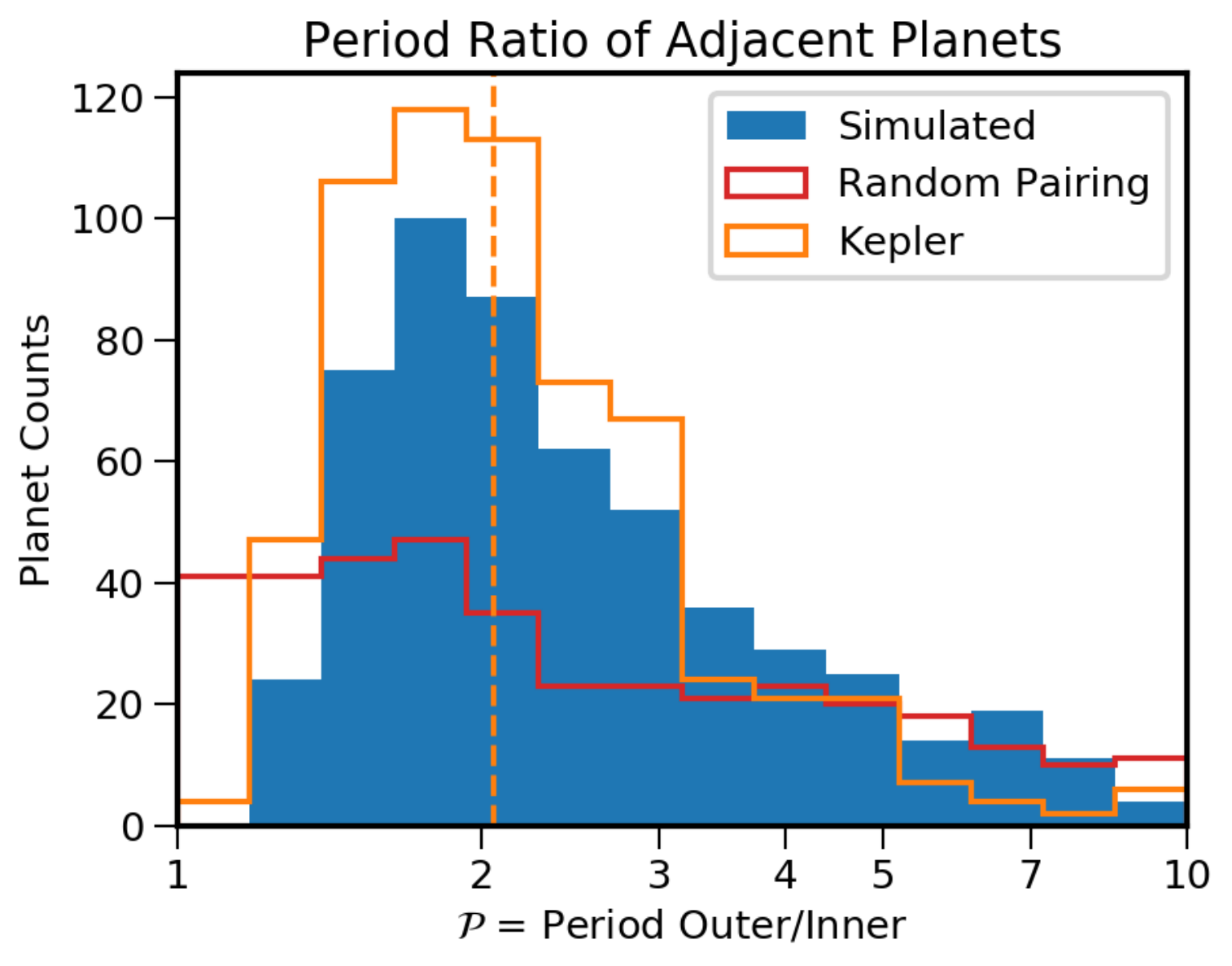}
        \caption{Simulated (blue) versus observed (orange) period ratio between adjacent planets in multi-planet systems. The red line shows, for comparison, a simulation where planet orbits are randomly drawn from the period-radius distribution (Figure \ref{f:panels}) instead of regularly spaced.
    }
    \label{f:obs:dP}
\end{figure}

\begin{figure} % Fig 13
    %\includegraphics[width=\linewidth]{\string~/EOStool/png/zoom_multi/output/{innerperiod}.png}
    % ./paper_zoom.py multi noextra
    % cp ~/EOStool/png/zoom_multi/output/innerperiod.png png/f_pinner_noextra.png
    \includegraphics[width=\linewidth]{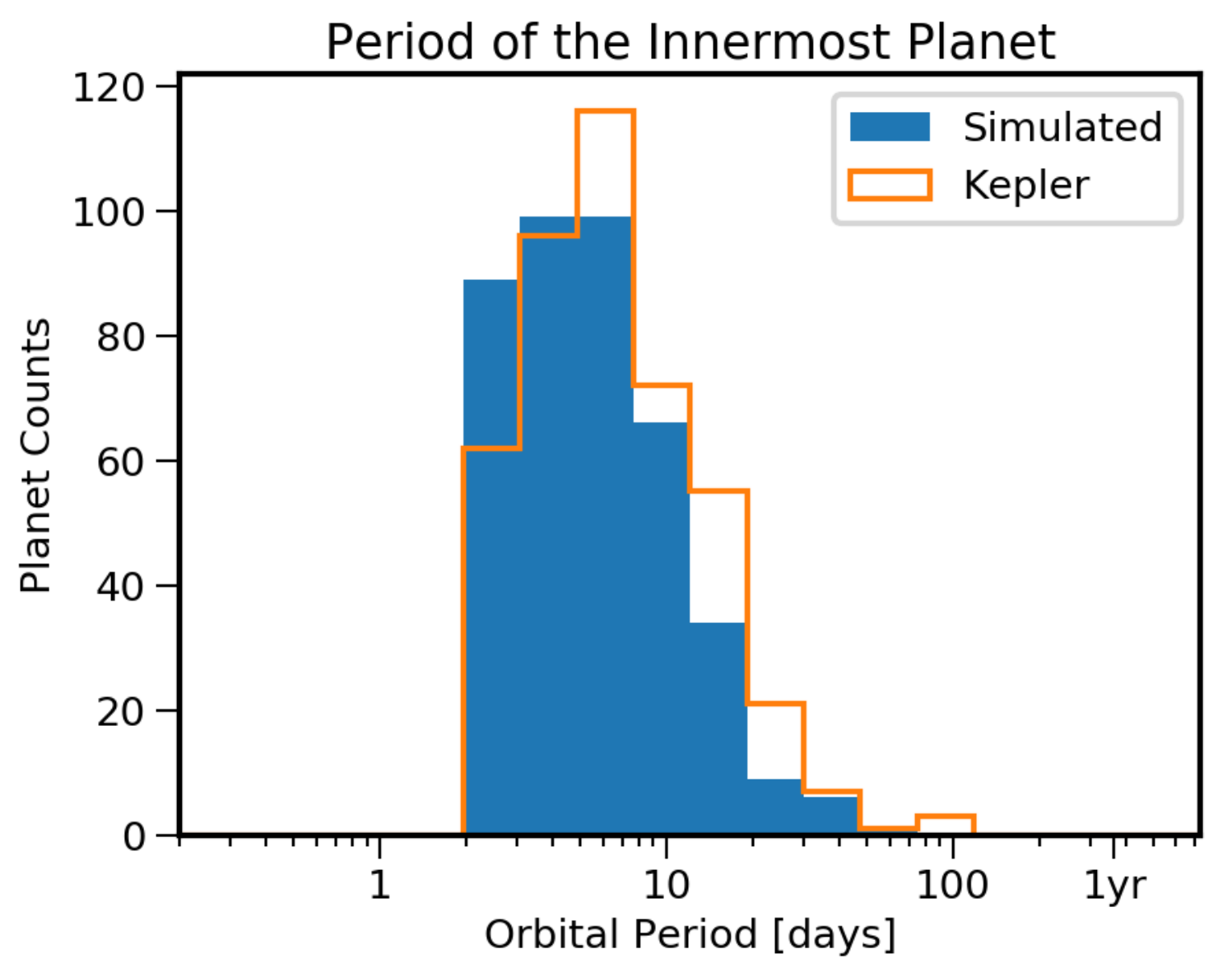}
                \caption{Simulated (blue) versus observed (orange) location of the innermost planet in each system. 
    }
    \label{f:obs:inner}
\end{figure}

We calculate three additional summary statistics for multi-planet systems: the frequency of multi-planet systems ($N_k$, Fig. \ref{f:multi:obs}); the period ratio of adjacent planet pairs ($\mathcal{P}$, Fig. \ref{f:obs:dP}); and the distribution of the locations of the innermost planet in each system ($P_\text{in}$, Fig. \ref{f:obs:inner}), all evaluated within $R=[0.5,6] R_\oplus$ and $P=[2,400]$. As discussed in the previous section, the observed planet populations are subject to detection biases, and 
a proper comparison requires comparing the observed distribution to the simulated distribution, which is what \epos does.
All summary statistics are calculated in the same way for observed planets (orange lines) and simulated planets (blue histograms) from the set of orbital periods and host star IDs.
We use 2-sample KS tests to calculate the probabilities that the distribution of inner planet orbital periods ($p_\text{in}$) and orbital period ratios ($p_\mathcal{P}$) are drawn from the same distribution as the observations. 
We use a Pearson $\chi^2$ test to calculate the probability, $p_{N}$, that the multi-planet frequencies of the simulated sample are drawn from the same distribution as the observations. 

The frequency distribution of planets in multi-planet systems is shown in Figure \ref{f:multi:obs} for a model with $m=7$ planets per system, a mode for the mutual inclination of $\Delta i=2\degr$, and $\fiso=0.4$.
The hatched region indicates simulated planets that are on isotropic orbits to match the excess of single-transiting systems.
We also show a model with only co-planar orbits ($i=0\degr$ and $\fiso=0$, green) that over-predicts the frequency of multi-planet systems, particularly at high numbers of planets per system. A model with planets on isotropic orbits ($\fiso=1$, red) under-predict the frequency of all multi-planet systems.

The period ratio of adjacent planet pairs is shown in Figure \ref{f:obs:dP}. The blue histogram shows the observable period ratios of the example model where planets are regularly spaced according to the period ratio distribution of Eq. \ref{eq:dP}. The shape of the period ratio distribution qualitatively reproduces the observed distribution (orange), with a peak near a period ratio of $\mathcal{P}\approx 2$ and a tail towards large orbital period ratios.
For comparison, the red line shows a simulation where planetary systems are not regularly spaced, which is constructed by randomly drawing $m=7$ planets from the period-radius distribution of Eq. \ref{eq:pdf} (Figure \ref{f:panels}). These simulated systems have an observable period ratio distribution that is much wider than observed, indicating that planetary systems are regularly spaced (See also \citealt{2018AJ....155...48W}).

The distribution of the location of the innermost detected planet in each multi-planet systems is shown in Figure \ref{f:obs:inner}. 
We only focus on the innermost planet of a detected multi-planet system, 
as for observed single-planet systems it can not be derived from the observable if they are intrinsically single or part of a multi-planet system with non-transiting planets. 
The distribution of innermost detected planets constrains the fraction of planetary systems without planets at short ($\lessapprox 20$ days) orbital periods.

\subsection{Step 5: Fitting Procedure}\label{s:mcmc}
With the framework to compare the parameterized distributions of exoplanets to observables we proceed to constrain the parameters to provide the best match to the observed planetary systems. The runtime of steps 1-4 is less than a tenth of a second in occurrence rate mode and less than a second in multi-planet mode, allowing for an comprehensive sampling of parameter space.  We use \emcee \citep{2013PASP..125..306F}, an open-source \python implementation of the algorithm by \cite{Goodman:2010et} to sample the parameter space and estimate the posterior distribution of parameters.

\subsubsection{Occurrence rate mode}
In occurrence rate mode, we explore the 7-dimensional parameter space consisting of the number of planets per star, $\eta$, the orbital period distribution, $P_\text{break}$, $a_P$, $b_P$, and the planet radius distribution, $R_\text{break}$, $a_R$, $b_R$. The summary statistics to evaluate the model given the observations are the number of detected planets, $N$; the planet orbital period distribution, $\{P\}$; and the planet radius distribution $\{R\}$. 
We use Fisher's method \citep{fisher1925statistical} to combine the probabilities from the summary statistic into a single parameter:
\begin{equation}
\mathcal{L}_\text{occ}= -2 (\ln(p_N)+\ln(p_P)+\ln(p_R))
\end{equation}
which we use as the likelihood of the model given the data in \emcee, e.g. $\mathcal{L} \propto \mathcal{L}(N,\{P\},\{R\}|\eta, P_\text{break}, a_P, b_P,$ $R_\text{break}, a_R, b_R)$.

\begin{figure} % Fig 14
	% cp ~/EOStool/png/dr25_vetting/mcmc/pdf.zoom.png png/f_single_mcmc.png
    \includegraphics[width=\linewidth]{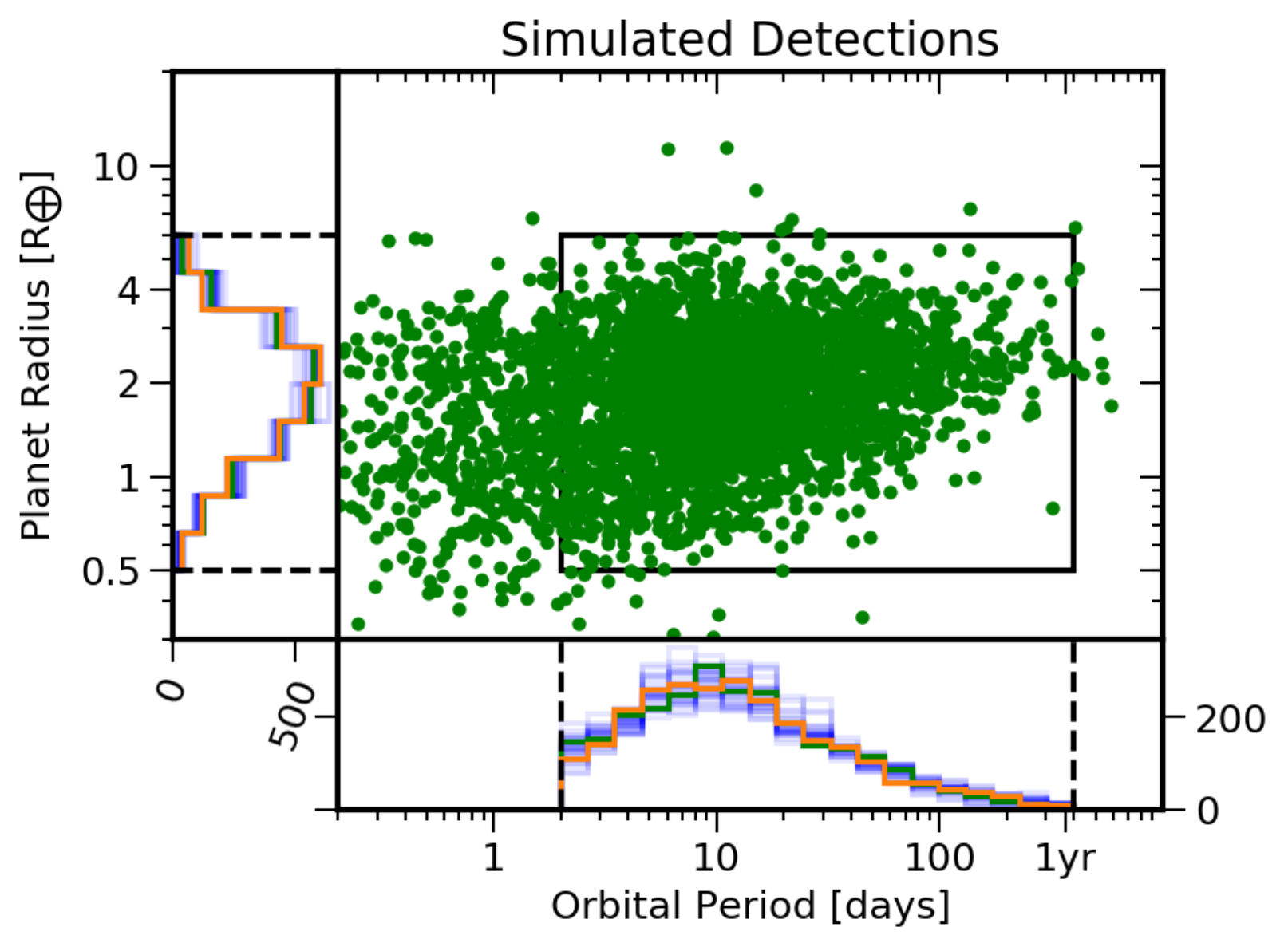}
    \caption{Period-radius distribution of detected planets in occurrence rate mode. The green population shows the population generated from the best-fit parameters. 
    The black box and dashed lines indicate the range of orbital periods and planet sizes that is included in the observational comparison.
    The side histograms show the marginalized simulated distributions compared with the observed populations in orange. The blue lines show 30 samples from the posterior.
    }
    \label{f:single:mcmc}
\end{figure}

%Best-fit values
%  pps= 4.94 +1.28 -1.17
%  P break= 11.8 +5.57 -3.09
%  P1= 1.54 +0.475 -0.261
%  P2= 0.26 +0.121 -0.2
%  R break= 3.33 +0.267 -0.372
%  R1= -0.467 +0.181 -0.186
%  R2= -5.93 +2.01 -2.58
\begin{table}
	\centering
	\title{Best-fit parameters}
   	\begin{tabular}{llll}
   	\hline\hline
	Parameter & Example & Best-fit \\
	\hline
	$\eta$				& $2.0$ 	& $4.9^{+1.3}_{-1.2}$ \\
	$P_\text{break}$(days)	& $10$	& $12^{+5}_{-3}$		\\
	$a_P$	 			& $1.5$ 	& $1.5^{+0.5}_{-0.3}$ \\
	$b_P$				& $0.0$  	& $0.3^{+0.1}_{-0.2}$ \\
	$R_\text{break}$($R_\oplus$)	& $3.0$ 	& $3.3^{+0.3}_{-0.4}$ \\
	$a_R$				& $0.0$ 	& $-0.5^{+0.2}_{-0.2}$	 \\
	$b_R$				& $-4$ 		& $-6^{+2}_{-3}$ \\	
   	\hline\hline
   	\end{tabular}
   	\caption{Fit parameters for the example model and the best-fit solutions with $1\sigma$ confidence intervals.}
	\label{t:fit:single}
\end{table}

We run \emcee with 200 walkers for 5,000 iterations, allowing for a 1000-step burn-in to reach convergence. For the initial positions of the walkers we use the parameters of the example model, see Table \ref{t:fit:single}. The MCMC chain and parameter covariances are shown in the Appendix (Figures \ref{f:chain} and \ref{f:corner}). The best-fit values and their $1\sigma$ confidence intervals are calculated as the $50\%$, and $16\%$ and $84\%$ percentiles, respectively. The simulated model for the best-fit parameters and for 30 samples of the posterior is shown in Figure \ref{f:single:mcmc}.

\begin{figure} % Fig 15
	% run w/ Rzoom=1.0 i.o. 0.5
	% cp ~/EOStool/png/dr25_plavchan_all/occurrence/posterior_x.png png/
	% cp ~/EOStool/png/dr25_plavchan_all/occurrence/posterior_y.png png/
	% cp ~/EOStool/png/dr25_vetting/occurrence/posterior_x.png png/
	% cp ~/EOStool/png/dr25_vetting/occurrence/posterior_y.png png/
    \includegraphics[width=\linewidth]{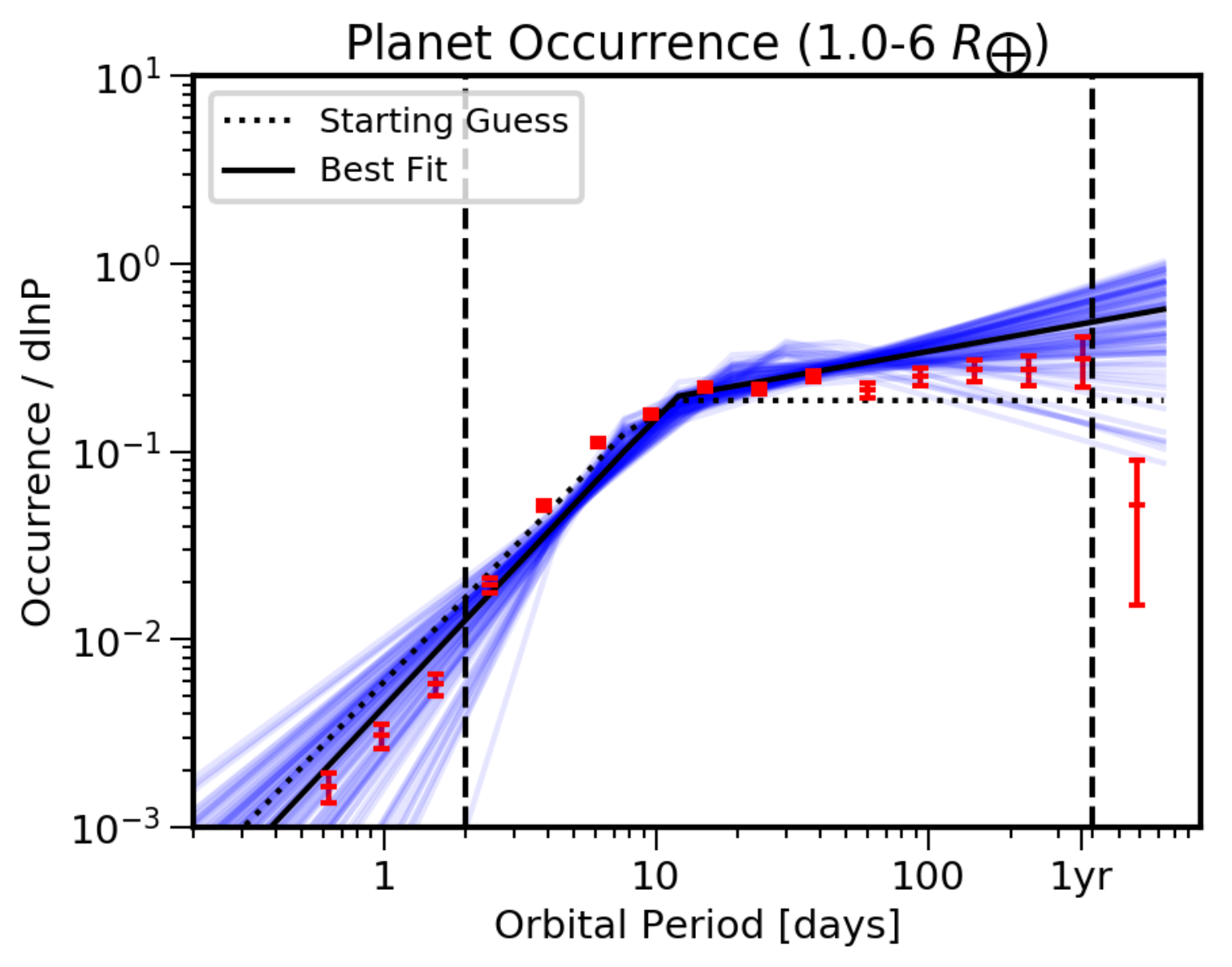}
    \includegraphics[width=\linewidth]{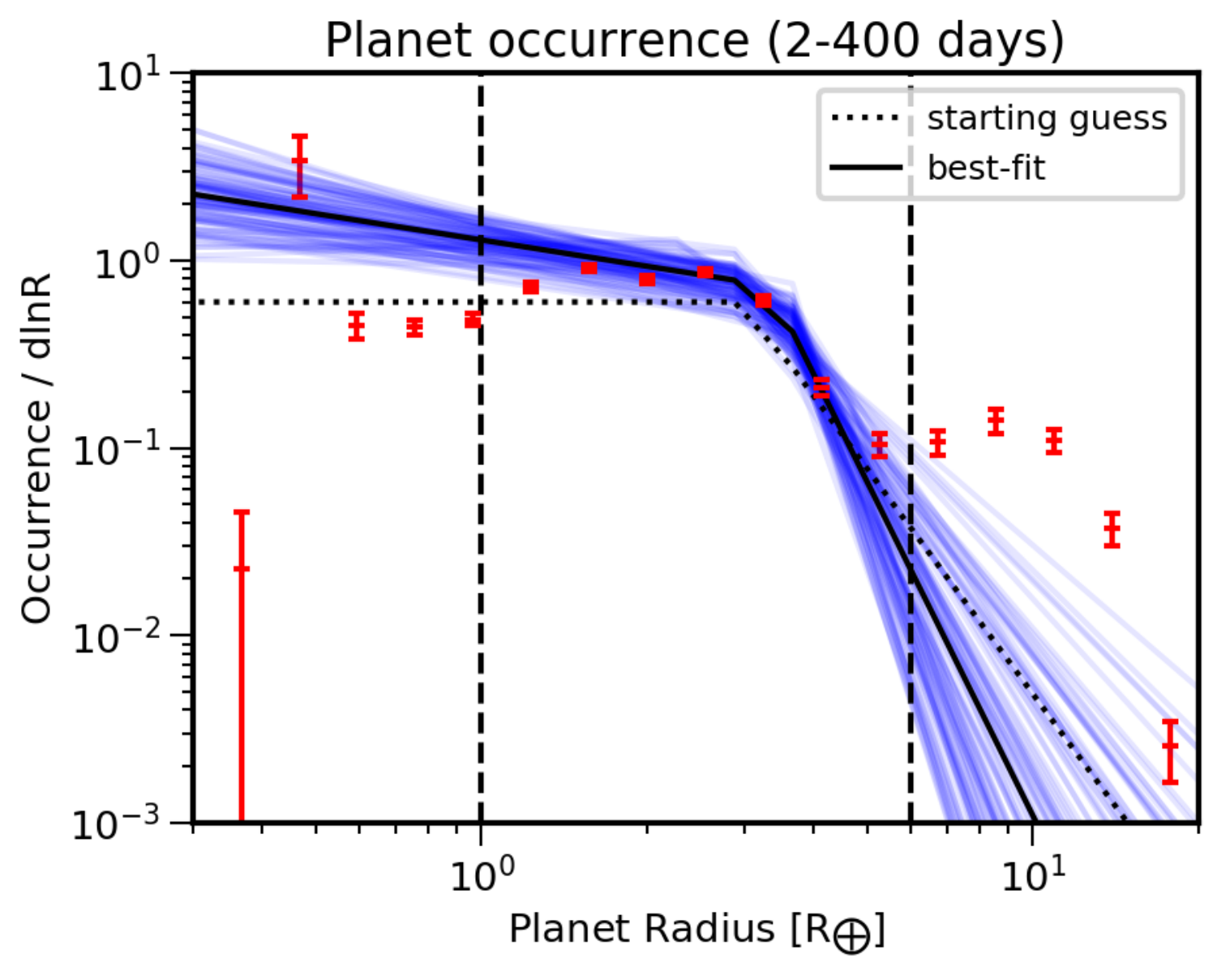}
    \caption{Posterior orbital period distribution (top) and planet radius distribution (bottom). The red bars show the occurrence rates estimated using the inverse detection efficiencies for comparison.
    Note that the occurrence rates underestimate the true distribution in bins that include regions where \kepler has not detected any planet candidates, in particular $R<1.5 R_\oplus$ and $P>50$ days (see Figure \ref{f:planets}).
    }
    \label{f:posterior}
\end{figure}

\begin{figure}% fig16
    %cp ~/EOStool/png/dr25_vetting/occurrence/bins.png png/f_planets.png
    \includegraphics[width=\linewidth]{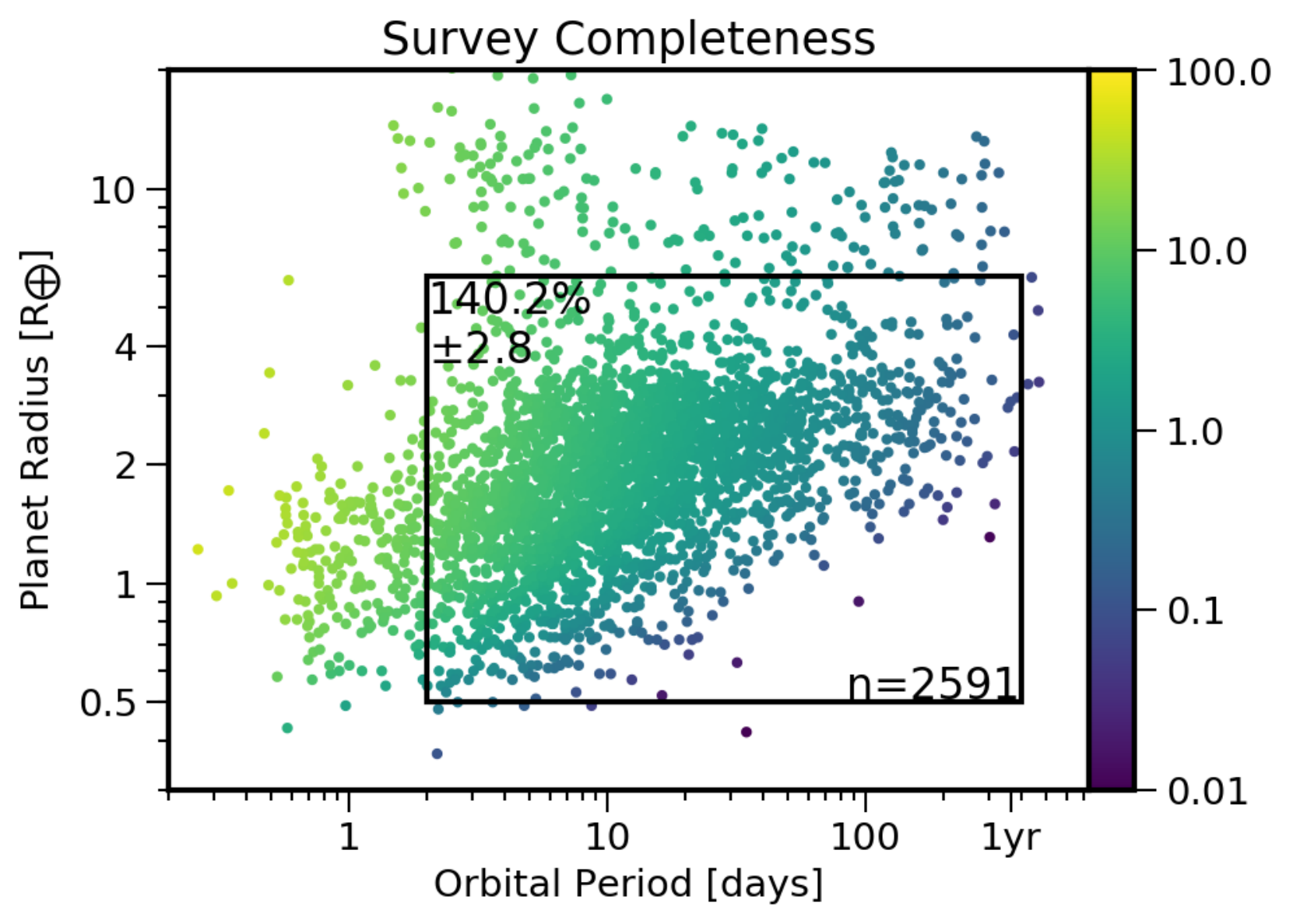}
    \caption{\kepler \texttt{DR25} candidate list, color coded by survey completeness. The sample includes only dwarf stars ($\log g ~\textless 4.2$) and planet candidates with a disposition score larger than 0.9. The planet occurrence rate estimated from Eq. \ref{eq:occ} is $\eta_\text{obs}=1.40\pm0.03$.
    }
    \label{f:planets}
\end{figure}

The posterior distribution of planet radius and orbital period are shown in Figure \ref{f:posterior}. As a sanity check, we calculate occurrence rates as function of planet radius and orbital period using the inverse detection efficiency method. The occurrence per bin is calculate as
\begin{equation}\label{eq:occ}
\eta_\text{bin}= \frac{1}{n_\star} \Sigma_j^{n_p} \frac{1}{\text{comp}_j}
\end{equation}
where ${\rm comp}_j$ is the survey completeness evaluated at the radius and orbital period of each planet in the bin (Figure \ref{f:planets}). The posterior distributions provide a decent match to the binned occurrence rates, with three notable deviations.  At $P \gtrsim 50$ days and $R \lesssim 1.4 R_\oplus$ occurrence rates are lower than the posterior. We attribute this to these bins including regions where \kepler has not detected planet candidates (see Figure \ref{f:planets}) and occurrence rates are therefore underestimated.
The broken power--law in planet radius does not describe the population of giant planets at $\sim 10 ~R_\oplus$, and we therefore in the following restrict our observational comparison to $\textless 6 ~R_\oplus$.

%  pps= 0.672 +0.166 -0.12
%  P break= 11.7 +2.85 -2.4
%  P1= 1.61 +0.363 -0.212
%  P2= -0.942 +0.408 -0.531
%  log D= -0.389 +0.0731 -0.052
%  sigma= 0.184 +0.0524 -0.0405
%  inc= 2.08 +1.01 -0.853
%  f_iso= 0.376 +0.0765 -0.0802

\begin{table}
	\title{Best-fit parameters}
	\centering
   	\begin{tabular}{lll}
   	\hline\hline
	Parameter & Example		& Best-fit \\
	\hline
	$\eta_s$ 			& $0.4$		& $0.67^{+0.17}_{-0.12}$ \\
	$P_\text{in}$ (days)& $10$		& $12^{+3}_{-2}$		\\
	$a_P$	 			& $1.5$ 	& $1.6^{+0.4}_{-0.2}$ \\
	$b_P$				& $-1$  	& $-0.9^{+0.4}_{-0.5}$ \\
	$R_\text{break}$ ($R_\oplus$)	& $3.0$	& $3.3$  \\
	$a_R$				& $0.0$		& $-0.5$ \\
	$b_R$				& $-4$ 		& $-6$ \\
	\hline
	$\log D$			& $0.4$		& $-0.39^{+0.07}_{-0.05}$ \\
	$\sigma$			& $0.2$		& $0.18^{+0.05}_{-0.04}$ \\
	\hline
	$m$ 				& $10$ 		& $10$ \\
	$\Delta i$ ($\degr$) & $2$		& $2.1^{+1.0}_{-0.8}$ \\
	\fiso				& $0.4$		& $0.38^{+0.08}_{-0.08}$ \\
%	\fcor				& $0.5$		&  \\
   	\hline\hline
   	\end{tabular}
   	\caption{Fit parameters for the example model in multi-planet mode and the best-fit solutions with $1\sigma$ confidence intervals. Parameters $R_\text{break}$, $a_R$, and $b_R$ were fixed to their best-fit solutions from occurrence rate mode. %\fcor was also fixed. 
	}
	\label{t:fit:multi}
\end{table}

\begin{figure*}
	% cp ~/EOStool/png/dr25_vetting_multi/mcmc/{pdf.zoom}.png png/f_mcmc_multi_a.png
	% cp ~/EOStool/png/dr25_vetting_multi/mcmc/multiplicity.png  png/f_mcmc_multi_b.png
	% cp ~/EOStool/png/dr25_vetting_multi/mcmc/periodratio.png  png/f_mcmc_multi_c.png
	% cp ~/EOStool/png/dr25_vetting_multi/mcmc/innerperiod.png  png/f_mcmc_multi_d.png
	\centering
    \includegraphics[width=0.45\linewidth]{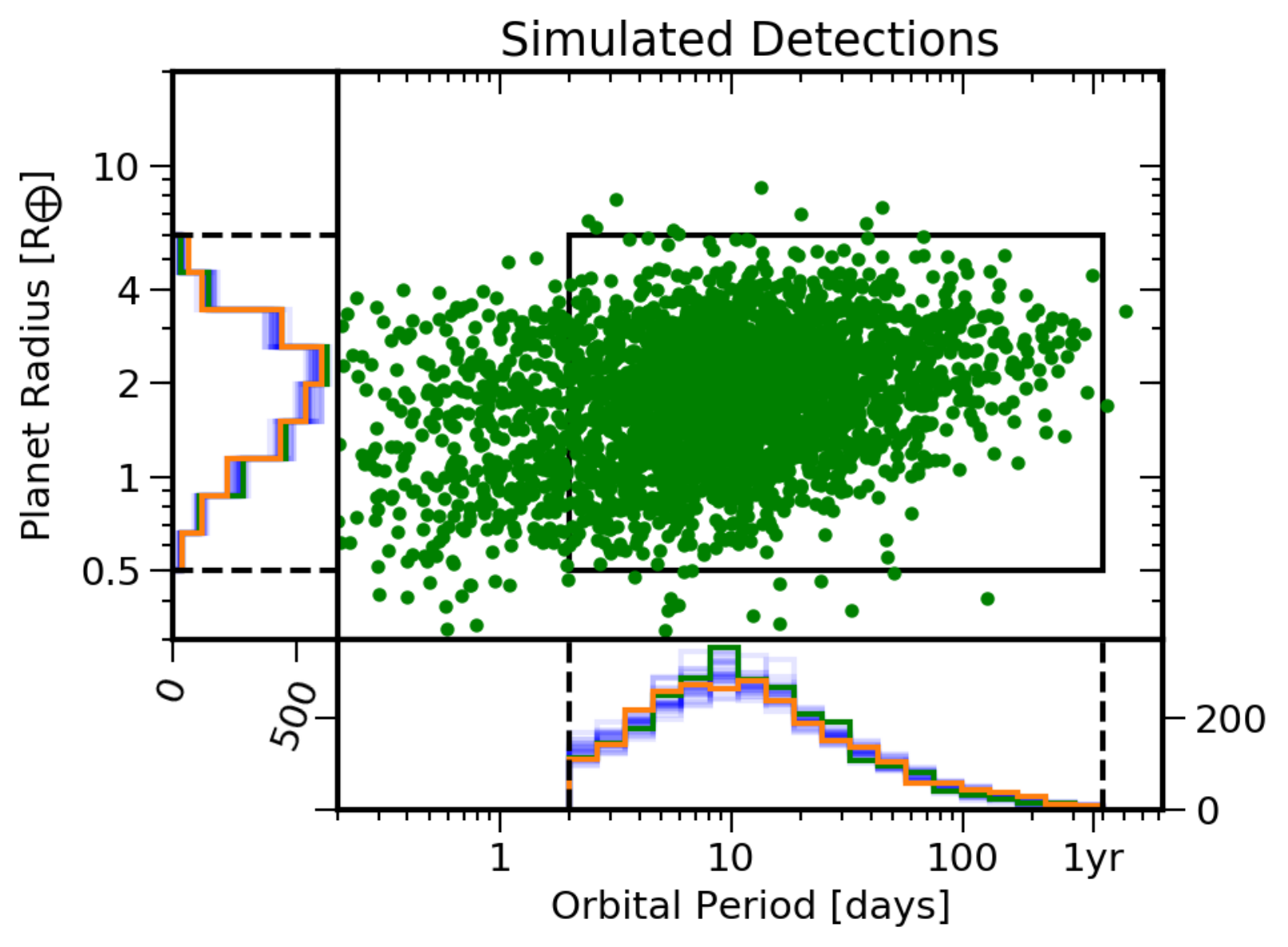}
    \includegraphics[width=0.45\linewidth]{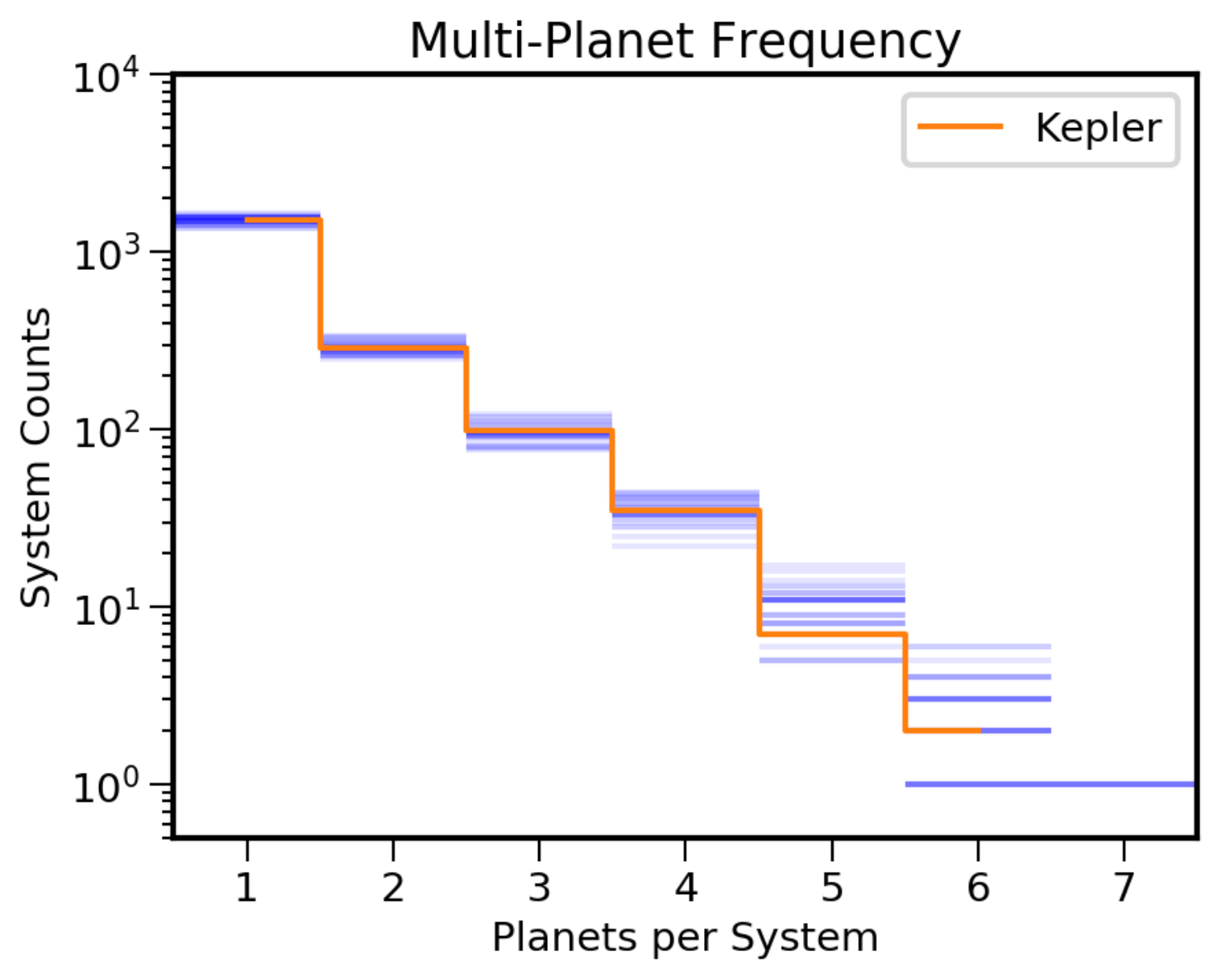}
	\includegraphics[width=0.45\linewidth]{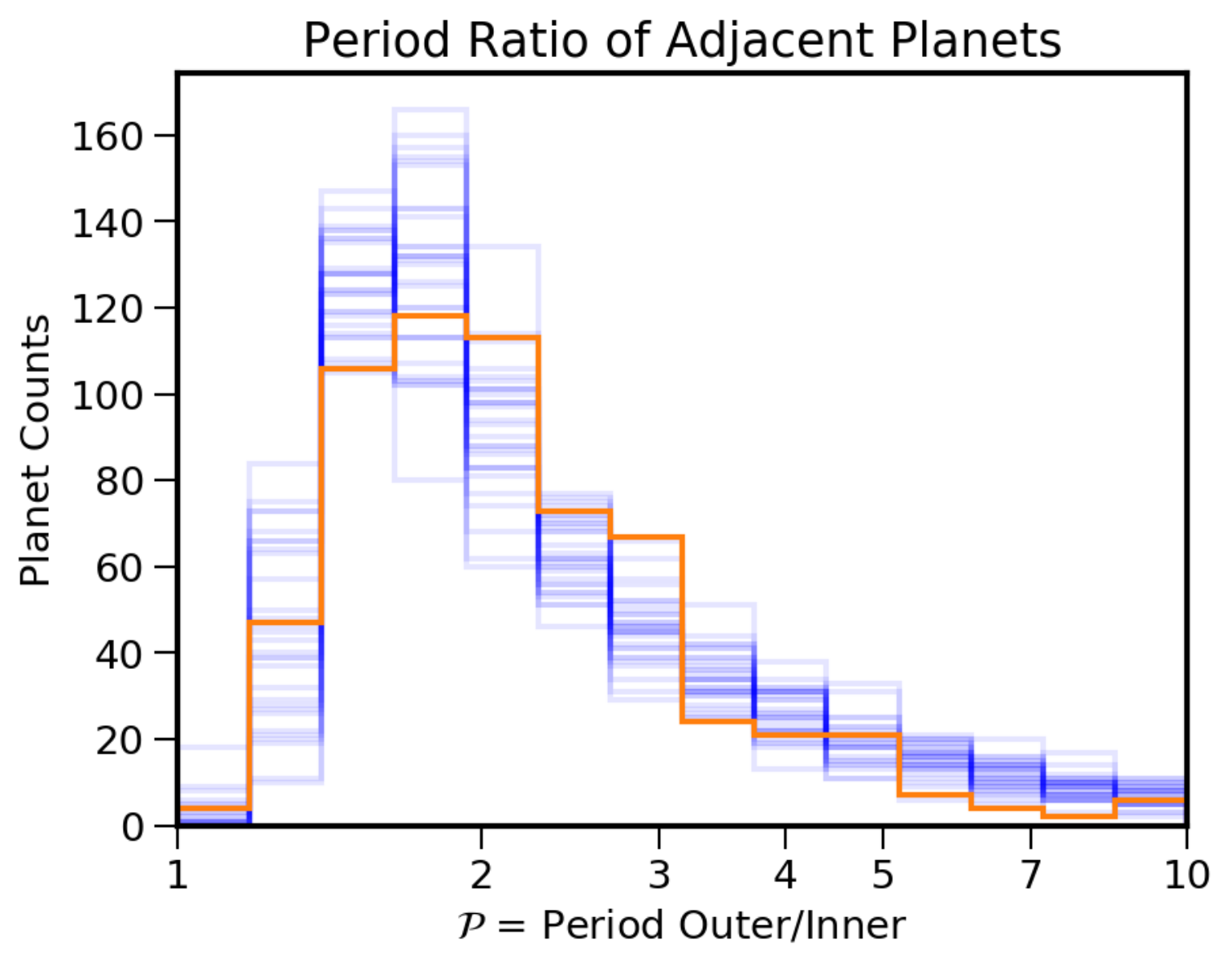}
	\includegraphics[width=0.45\linewidth]{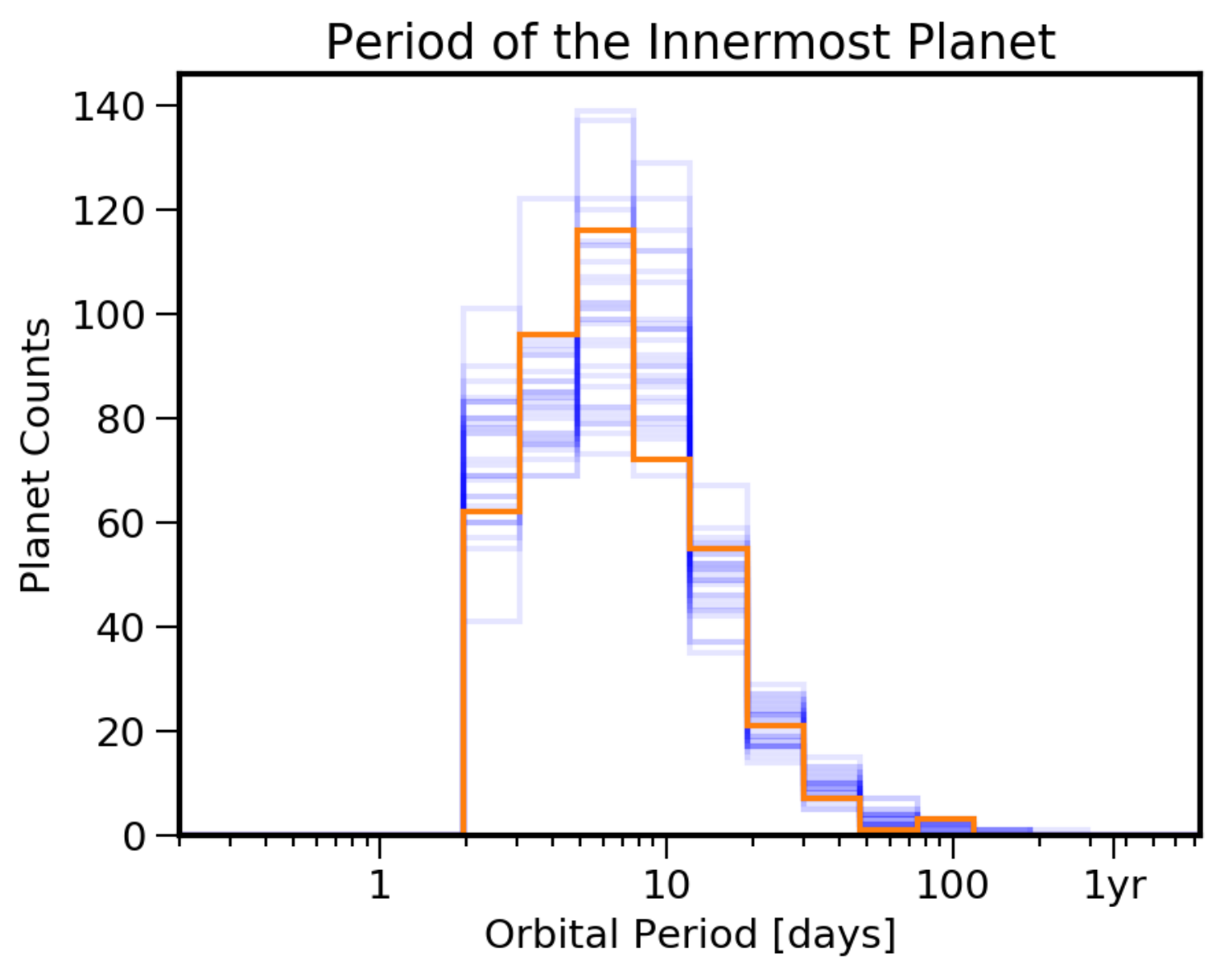}
    \caption{Posterior predictive plots of simulated planetary systems. 30 samples from the posterior are shown in blue and the observed systems are shown in orange. The top left panel show the best-fit period-radius distribution in green, see Figure \ref{f:single:mcmc}. 
    The black box and dashed lines indicate the range of orbital periods and planet sizes that is included in the observational comparison.
    The top right panel shows the frequency distribution of multi-planet systems with $k$ planets. The bottom left panel show the period ratio distribution of adjacent planets. The bottom right panel shows the orbital period distribution of the innermost planet in each multi-planet system.
    }
    \label{f:mcmc:multi}
\end{figure*}

\subsection{Multi-planet systems}
In multi-planet mode, we explore the 9-dimensional parameter space consisting of: 
the fraction of stars with  planetary systems, $\eta_s$;  
the mode of the mutual inclination distribution, $\Delta i$; 
the orbital period distribution of the inner planet, $P_\text{break}$, $a_P$, $b_P$; 
the period ratio distribution, $D$ and $\sigma$;
and the fraction of isotropic systems, $\fiso$. 
The summary statistics to evaluate the model given the observations are the frequency of multi-planet systems, $\{N_k\}$; the orbital period distribution, $\{P\}$; the planet orbital period ratio distribution, $\{\mathcal{P}\}$; the distribution of the innermost planet in the system,$\{P_\text{in}\}$. 
We use Fisher's method to combine the probabilities from the summary statistic into a single parameter:
\begin{equation}
\mathcal{L}_\text{multi}= -2 (\ln(p_N)+ \ln(p_{N,k})+\ln(p_{P})+\ln(p_\mathcal{P})+ \ln(p_\text{in}))
\end{equation}
which we use as the likelihood of the model given the data in \emcee, e.g., $\mathcal{L} \propto \mathcal{L}(N_k,\{P\},\{\mathcal{P}\}, \{P_\text{in}\}|\eta, P_\text{break},$ $a_P, b_P, D, \sigma, \Delta i, \fiso)$.

The planet radius distribution is fixed to minimize the amount of free parameters, and we use the best-fit values constrained in the previous section and listed in Table \ref{t:fit:multi}. The number of planets per system, $m$, is fixed to $10$ because it is not well constrained in the fitting: Systems with fewer than 6 planets do not reproduce the observed multiplicity distribution as there are two detected sextuple systems ($N_6=2$). 
Systems with more than 7 planets have the same observation signature as the 8th, 9th, etc., planet in the system will typically remain undetected, as the combined likelihood of detecting all planets in such systems is too small to allow detections given the size of the Kepler sample.

We sample the parameter space with \emcee using 100 walkers for 2,000 iterations, allowing for a 500-step burn-in. The MCMC chain and parameter covariances are shown in the appendix (Figure \ref{f:multi:corner}). The simulated observables of the best-fit models and of 30 samples from the posterior are shown in Figure \ref{f:mcmc:multi}. We will discuss the results in the next section.

\section{Results}\label{s:results}
By using \epos in parametric mode find that the planet occurrence rate of the \kepler mission are well described by a broken power--law in orbital period and planet radius in the region $2 < P < 400$ days and $0.5 < R_\oplus < 6$. We estimate the planet occurrence rate, the \textit{average} number of planets per star, to be $\eta= 2.4^{+0.5}_{-0.5}$ in this regime and $\eta=4.9^{+1.3}_{-1.2}$ in the simulated range ($0.4 < P < 730$ days and $0.3 < R_\oplus < 20$). This number is higher than the occurrence rate calculated from inverse detection efficiencies ($\eta_\text{inv}= 1.40\pm0.03$), which underestimate the occurrence rates for small planets at long orbital periods.
These results are largely consistent with previous occurrence rate studies collected in the SAG13\footnote{\url{https://exoplanets.nasa.gov/system/internal_resources/details/original/680_SAG13_closeout_8.3.17.pdf}} literature study on planet occurrence rates (see \citealt{2018arXiv180209602K} for details).

The observed population of exoplanets is well described by a population of regularly-spaced planetary systems with 7 or more planets that orbits $\eta_s=67^{+17}_{-12}\%$ of stars. 
We find that the mutual inclinations are consistent with the Kepler dichotomy: Half the planet population is in planetary systems that have nearly co-planar orbits, here described by a Rayleigh distribution with $\Delta i=2.1^{+1}_{-0.8} \degr$, consistent with previous estimates that find ranges between $1-3\degr$ \citep{2011ApJS..197....8L,2012ApJ...761...92F,2014ApJ...790..146F,2016ApJ...816...66B}. The other half of planets appear as single-planet transiting systems, which we model as multi-planet systems with isotropically distributed orbital inclinations. 
The typical orbital period ratio between adjacent planets is $\mathcal{P}=1.8$ but with a wide distribution, consistent with previous analysis of spacings between planets \citep{2013ApJ...767..115F}.

\begin{figure}
	% cp ~/EOStool/png/dr25_vetting_multi/mcmc/posterior.linear_x.png png/f_posterior_inner.png
        \includegraphics[width=\linewidth]{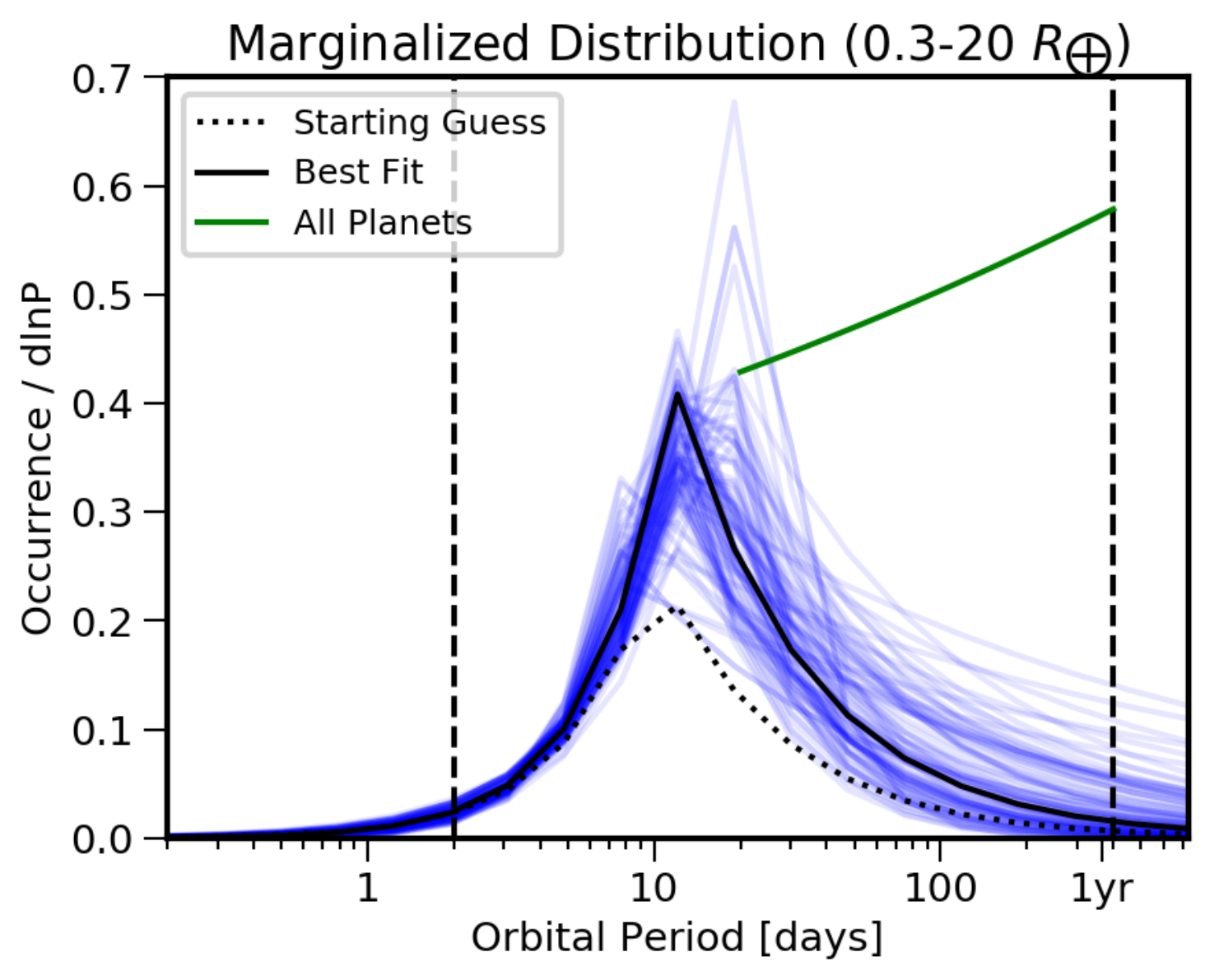}
    \caption{Marginalized orbital period distribution of the innermost planet in the system.
    The green line indicates the distribution of all planets in the system.
    }
    \label{f:posterior:inner}
\end{figure}

We discover that the inner edge of planetary systems are clustered around an orbital period of 10 days (Fig. \ref{f:posterior:inner}), reminiscent of the protoplanetary disk inner edge which is located at $\sim 0.1$ au for pre-main-sequence sun-like stars \citep{2007prpl.conf..539M}. While the break in planet occurrence rate around 10 days has been previously connected to the disk inner edge \citep{2015ApJ...798..112M,2017ApJ...842...40L}, this is the first time a peak in the occurrence rate distribution of sub-Neptunes has been identified. The posterior distribution of the innermost planet peaks at an orbital period of $12\pm2$ days and decays towards shorter and longer orbital period with a power--law index of $a_P=1.6^{+0.4}_{-0.2}$ and $b_P=-0.9^{+0.4}_{-0.5}$, respectively. 
The decay towards long orbital periods is surprising, since planet occurrence rates are constant or slightly increasing in this range (green line). Planets exterior to the break are therefore mostly the 2nd, 3rd, etc., planet in the system. 
We also show that systems with inner planets at orbital periods of $~100$ days are intrinsically rare, and we discuss the implications for planet formation theories and the origins of the solar system below.

\section{Discussion}

\subsection{How rare is the solar system?}
Our modeling analysis indicates that multi-planet systems without planets interior to $P_\text{in} \sim 100$ days are intrinsically rare. We estimate, based on parametric distributions of planet parameters, that $8^{+10}_{-5}\%$ of planetary systems have no planet interior to the orbit of Mercury and $3^{+5}_{-2}\%$ have no planet interior to Venus. This implies that solar system may simply be in the tail of the distribution of exoplanet systems.

On the other hand, this comparison with the solar system is made under the assumption that planet orbital architectures (inclinations, spacings, inner planet location) are independent of radius and orbital period. This assumption will need to be verified with additional data. The orbital architectures of the \textit{Kepler} planetary systems are most constrained by planets that are larger and closer in, typically $P\sim 10$ days and $R\sim 2 ~R_\oplus$ (see Figure \ref{f:single:mcmc}). A solar system analogue, if detected, would most likely appear as a single transiting systems in the \kepler data due the low probability that multiple terrestrial planets transit \citep[e.g.][]{2016ApJ...821...47B}. Hence, we can not rule out that the solar system may be part of a population of planetary systems with different orbital characteristics than those detected by the \kepler mission. 
 
We make a direct estimate of how many planetary systems without planets interior to Mercury or Venus could be present in the Kepler data, without extrapolating the distribution of orbital periods from closer-in planets. We generate a population of planetary systems with similar orbital properties as the solar system where we place the innermost planet at the orbital period of mercury (or Venus) but otherwise keep the same parameters as in the best-fit model. Such a simulated planet population with $\eta_s=67\%$ and $\fiso=0$ matches planet occurrence rates exterior to $P=50$ days for $P_\text{in}=88$ days. The simulated observations predict $~30$ multi-planet systems with an innermost planet exterior to $P=50$ days, while only $4$ are observed. This indicates that if a population of planetary systems without planets interior to Mercury existed it would be detectable in the Kepler data. We do note that most detectable planets in this range are larger than one earth radius, so the data do not directly constrain true solar system analogues. 
Using a mixture of co-planar systems and highly inclined systems (which appear as intrinsically single), we can rule out that more than $10\%$ of systems ($\fiso=0.9$) have $P_\text{in}=88$ days and $\Delta i=2$.

We repeat this exercise for systems where the innermost planet shares the orbital period of Venus ($P=225$ days). None of these simulated systems would be detectable as multi-planet systems, and no planetary systems with an orbital period larger than $150$ days are detected with Kepler. Hence we can not rule out that the Kepler exoplanet population contains a significant population of multi-planet systems without planets interior to Venus, though we do not find evidence that such a population exists.

\begin{figure} % fig 19
 	% cp ~/EOStool/png/solarsystem.letters.png png/
       \includegraphics[width=\linewidth]{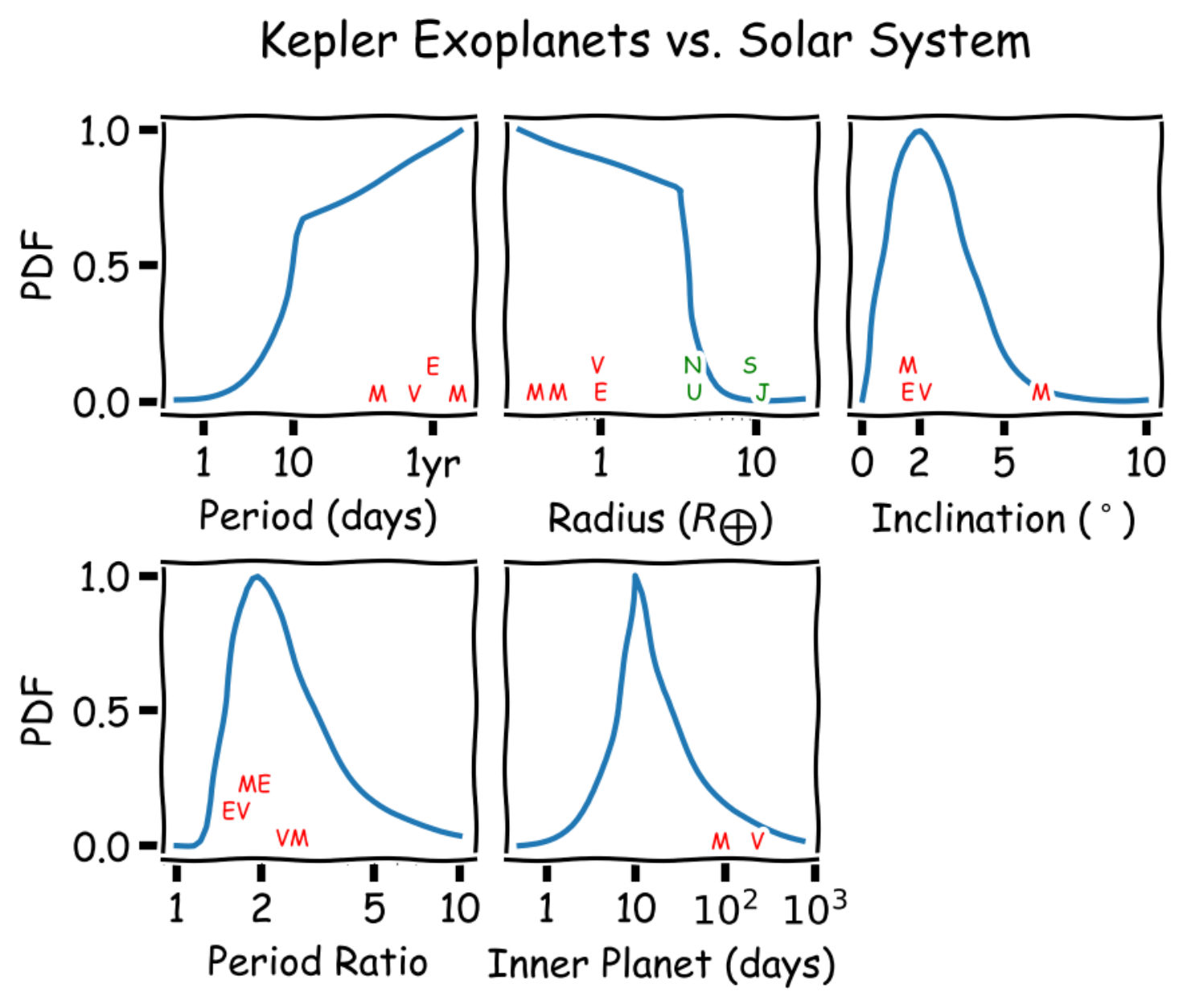}
    \caption{Best-fit distribution of planetary system properties from Kepler (blue) compared to the solar system terrestrial planets (red letters). The inclinations of the terrestrial planets are with respect to the invariable plane. 
    The orbital periods, radii, inclinations, and period ratios of the terrestrial planets lie near the peak of the distribution of kepler systems. The only notable exception is the orbital period of the innermost planet, where Mercury (93th percentile) and Venus (97th percentile) lie in the tail of the distribution. 
    }
    \label{f:ss}
\end{figure}

Overall, the solar system seems to be a typical planetary system in most diagnostics of orbital architectures (planetary radii, periods, period ratios, inclinations, and planets per system, Figure \ref{f:ss}). The only clear difference is in the period of the innermost planet where the solar system is an outlier. 
While the lack of super-earths/mini-Neptunes in the solar system is also notable \citep{2015ApJ...810..105M}, earth-sized planets are not intrinsically rare, and the size of the terrestrial planets does not make the solar system an outlier in the exoplanet distribution.

\subsection{Systems with habitable zone planets}
We also estimate $\eta_\oplus$, the number of earth-sized planets with earth-like orbital periods (``habitable zone'' planets), here defined as $0.9 P_\oplus < P < 2.2 P_\oplus$ and $0.7 R_\oplus < R < 1.5 R_\oplus$ based on the \cite{Kopparapu:2013fu} conservative habitable zone for a sun-like star that is representative of the most common star in the \kepler sample. By integrating the posterior distribution over the radius and orbital period range we find $\eta_\oplus= 36^{+14}_{-14}\%$ or $\Gamma_\oplus= \frac{\eta_\oplus}{dlnP ~dlnR}= 53^{+20}_{-21}\%$. These results are consistent with the estimate of $\Gamma_\oplus=60\%$ for GK dwarfs by \cite{2015ApJ...809....8B} though we note that there is a large dispersion in the literature\footnote{\url{https://exoplanets.nasa.gov/system/internal_resources/details/original/680_SAG13_closeout_8.3.17.pdf}}
on habitable zone planet occurrence rates based on the adopted completeness correction, the method of extrapolation into the habitable zone, and the planet sample selection. 
The confidence intervals on $\eta_\oplus$ include counting statistics and systematic uncertainties in extrapolating the planet radius and orbital period distribution out to the habitable zone. They do not include systematic uncertainties on the adopted stellar parameters, in particular the stellar radii which directly impacts the planet radii. For example, large uncertainties in the stellar radius of unresolved stellar binaries lead to over-estimating the occurrence of small planets \citep{2015ApJ...805...16C,2015ApJ...799..180S}.
Better observational constraints on the stellar properties are expected based on parallax measurments from the ESO \textit{Gaia} mission.
As a consistency check, we have repeated all calculations in this paper with the improved stellar radii and giant star classification from \citep{2018arXiv180500231B}. The best-fit parameteric distributions are consistent with those in table \ref{t:fit:single} within errors, and the habitable zone planet occurrence rate of $\eta_\oplus= 40^{+14}_{-14}\%$ is not significantly different.

The habitable zone planet occurrence rates we estimate is consistent with that of \cite{2015ApJ...809....8B} based on the Q1-Q16 catalog, but about a factor four higher than of the earliest estimate based on \kepler data from \cite{2013PNAS..11019273P}, which we attribute to an improved understanding of detection efficiency of \kepler.
This rate is comparable to that of M dwarfs estimated with inverse detection efficiency methods \citep{2015ApJ...807...45D}, though we leave a comparison between M dwarfs and FGK stars with consistent methodology for a future paper. 

The majority of simulated planets at long orbital periods ($P\sim 100$ days) are not intrinsically single, but are part of multi-planet systems where the innermost planets are typically at an orbital period of $P\sim10$ days. While the multi-planet statistics do not directly constrain planetary systems of earth-sized planets in the habitable zone of sun-like stars, the presence of planets on orbital periods of 10 days does provide a way to identify targets for future direct imaging missions, such as the HabEx and LUVOIR mission concepts. 

\cite{2017MNRAS.465.3495K} have shown that the presence of transiting planets at short orbital periods increases the chance of finding transiting planets at larger orbital periods. Our analysis indicates that, because most planets in multi-planet systems are not transiting, the probability of finding \textit{any} planet (transiting or non-transiting) is even higher. 
For example, the detection of an earth-mass planets at an orbital period of ten days with radial velocity measurements indicates a higher probability of finding a planet in the habitable zone. Depending on whether single-transiting planets are intrinsically single or highly-inclined multiples, $50-100\%$ of systems with close-in planets may also have planets in the habitable zone.

\subsection{Protoplanetary disk inner edges}
The peak in the location of the innermost planet in the system points to a preferred location of planet formation/migration in protoplanetary disks. This location may reflect either the inner edge of the region where planets form \citep{2013MNRAS.431.3444C,2015ApJ...798..112M,2017ApJ...842...40L} or that of a planet trap where planet migration stalls \citep{2007ApJ...654.1110T}. While the break in planet occurrence at $\sim 10$ days has been attributed to the disk inner edge before \citep{2015ApJ...798..112M,2017ApJ...842...40L}, the peak in the location of the innermost planet presents solid evidence that there is a preferred location for planet formation at 0.1 au, rather than a continuum of location between 0.1-1 au. In particular, this result is inconsistent with a wide regions between 0.1 and 1 au acting as a trap for individual planets \citep[e.g.][]{2014A&A...567A.121D}, and is more reminiscent of inward migration of multiple planets where the first planet is trapped at the inner disk edge and halts the migration of other planets \citep{2014A&A...569A..56C,2016MNRAS.457.2480C,2017MNRAS.470.1750I}.

\begin{figure*} % Fig 20
    \includegraphics[width=\linewidth]{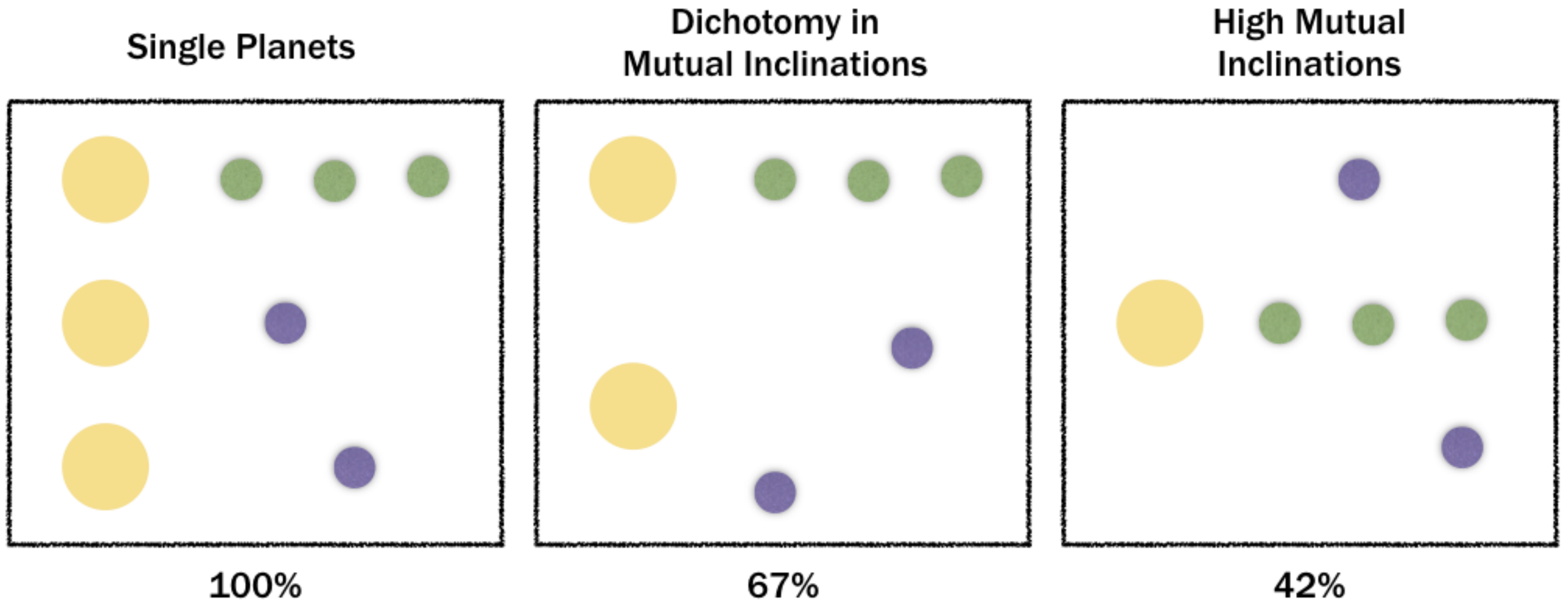}
    \caption{Illustration of how the estimated fraction of stars with planetary systems depends on the distribution of planets across stars. Nearly-coplanar systems orbit $31\%$ of stars (green) while the distribution of observed single transiting planets (purple) is less well constrained. 
    }
    \label{f:dichotomy}
\end{figure*}

\subsection{Do most stars have planets?}
Our analysis confirms that the \textit{average} number of planets per (sun-like) star is larger than one (\S \ref{s:para}). However, this does not imply that most stars have planets, since planets can be unevenly distributed among stars.
In (\S \ref{s:para}) we show that $\eta_s=67^{+17}_{-12}\%$ of stars can host multi-planet systems. An important assumption we made is that the apparent excess of single transits is because $\fiso=38\%$ of multi-planet systems have high mutual inclinations. The inclination distribution of exoplanets can not be uniquely constrained from transit survey data \citep{2012AJ....143...94T}. Different assumptions on the mutual inclination distribution lead to different estimates of the fraction of stars with planets, illustrated in Figure \ref{f:dichotomy}.

On the conservative end, the highly-inclined planets in otherwise coplanar multi-planet systems could contribute to the observed number of single transits (Figure \ref{f:dichotomy}, right panel). In this case, the dichotomy is an artifact of the assumption that mutual inclinations follow a Rayleigh distribution. A distribution with a larger tail towards higher inclinations  
could possibly fit the multi-planet statistics with a single population. In this case, the fraction of stars with planets drops to $\sim 41\%$. Alternatively, a population of intrinsically single planets could contribute to the observed single transiting systems (Figure \ref{f:dichotomy}, right panel). In this case, the estimated fraction of stars with planets increases to unity.

A way of discriminating between both scenarios may be the use of stellar obliquity. Transiting planets with high mutual inclinations or isotropic distributions have large obliquities with respect to the line of sight. There are indications that the stellar obliquity is larger for single-transit systems \citep{2014ApJ...796...47M}, indicating that they may have highly inclined orbits.
Several mechanisms have been proposed to disrupt initially co-planar systems, reducing their multiplicity and/or increasing their mutual inclinations, and giving rise to the population of single transiting systems. 
These mechanisms include dynamical instabilities within systems \citep{2015ApJ...806L..26V,2015ApJ...807...44P},
external perturbations from giant planets or stars \citep{2012ApJ...758...39J,2017AJ....153...42L,2017MNRAS.467.1531H,2017MNRAS.470.1750I,2018MNRAS.474.5114C} and host-star misalignment \citep{2016ApJ...830....5S}.
Alternatively, planets may form as intrinsically single systems in half the cases \citep{2016ApJ...832...34M}. Additional constraints on the single planet population are needed to constrain these scenarios. However, we stress that roughly half of the observed single transiting systems are part of multi-planet systems and have non-transiting or non-detected planets in the system, and hence the sample of ``singles'' is diluted with multi-planets, weakening any potential trends.  
Transit timing variations can also be used to detect additional non-transiting planets in the system \citep{2012Sci...336.1133N}.

The \kepler mission only detects planets out to $\sim1$ au and down to $\sim0.5 R_\oplus$. Hence, any estimate of the fraction of stars with planets is likely to be a lower limit. We do not see a clear indication that planet occurrence rate decreases towards the detection limits of Kepler. While planet occurrence rates calculated using the inverse detection efficiency seem to decrease for earth-size and smaller planets (Fig \ref{f:posterior}), this is also the region where the \kepler exoplanet list becomes incomplete. The posterior planet size distribution does not show this trend and is almost flat in log $R_p$, indicating that planets below the detection efficiency of \kepler could be extremely common. Since planetary systems tend to have planets of similar size \citep{2017ApJ...849L..33M,2018AJ....155...48W}, planetary systems with only planets smaller than the detection limit would be missed. Extending the planet size distribution down to the size of Ceres ($0.07 ~R_\oplus$) would increase the fraction of stars with planets to $100\%$.

We find that the orbital period distribution of sub-Neptunes is nearly flat in logarithm of orbital period. However, planets at long orbital periods are mostly members of systems with planets also at shorter orbital periods: there is no evidence for a large population of planetary systems with only long-period planets. Extrapolating the distribution of planetary systems to orbital periods larger than a year would add planets to existing systems, increasing the number of planets per star but not the fraction of stars with planets. Of course, such an extrapolation does not take into account the presence of additional populations of planets, which could be important, for example, if planets form more frequently at the snow line (exterior to $1$ au). Giant planets orbit $10-20\%$ of sun-like stars \citep{2008PASP..120..531C}, most of them beyond 1 au.
Microlensing surveys hint at the existence of a population of Neptune-mass planets around a significant fraction of M dwarfs \citep{2012Natur.481..167C,2016ApJ...833..145S}. Since these populations remain mostly undetected in the Kepler survey, it is not clear if they belong to the same stars, though there are some indications that they do \citep{2016ApJ...825...98H}.

\section{Summary}
We present the Exoplanet Population Observation Simulator, \epos, a \texttt{Python} code to constrain the properties of exoplanet populations from biased survey data. We showcase how planet occurrence rates and orbital architectures can be constrained from the latest \kepler data release, \texttt{DR25}. We find that:
\begin{itemize}
\item The \kepler exoplanet population between orbital periods of $2-400$ days and planet radii of $0.5-6 R_\oplus$ are well described by broken power--laws. The planet occurrence rate in this regime is $\eta=2.4\pm0.5$ consistent with previous works. The estimated planet occurrence rate in the habitable zone is $\eta_\oplus= 36\pm14\%$.
\item The observed multi-planet frequencies are consistent with ensemble populations that are a mix of nearly co-planar systems and a population of planets whose orbital architectures are unconstrained, consistent with the previously reported Kepler dichotomy. $62\pm8 \%$ of exoplanets are in systems with 6 or more planets, orbiting $42\%$ of sun-like stars. The remaining $38\pm8 \%$ of exoplanets could be intrinsically single planets or be part of multi-planet systems with high mutual inclinations, raising the fraction of stars with planets to somewhere between $45\%$ and $100\%$.
\item The mutual inclinations of planetary orbits can be described by a Rayleigh distribution with a mode of $i=2\pm1\degr$. The spacing between adjacent planets follows a wide distribution with a peak at an orbital period ratio $\mathcal{P}=1.8$, though detection biases shift the peak of the observed distribution to shorter orbital period ratios. 
\item The distribution of the innermost planet in the system peaks at an orbital period of $P\approx 10$ days. Planetary systems without planets interior to Mercury's and Venus' orbit are rare at $8\%$ and $3\%$, respectively.  
\end{itemize}

The \epos code presented in this paper provides a first step in a larger effort to constrain planetary orbital architectures from exoplanet populations. The next step will be to use more complex models of planetary system architectures based on planet formation models. Future directions include incorporating additional exoplanet survey data from radial velocity, microlensing, and direct imaging, as well as from transit surveys such as K2 and TESS.

\acknowledgments
We thank an anonymous referee for a constructive review of the manuscript. 
We also thank Shannon Dulz and Peter Plavchan for feedback on an early manuscript draft, Ed Bedrick for advice on statistical methods, and Daniel Huber for advice regarding stellar properties. We acknowledge helpful conversations with Carsten Dominik, Eric Ford, Daniel Carrera, Renu Malhotra, and Rachel Fernandes.
This material is based upon work supported by the National Aeronautics and Space Administration under Agreement No. NNX15AD94G for the program “Earths in Other Solar Systems”. The results reported herein benefited from collaborations and/or information exchange within NASA’s Nexus for Exoplanet System Science (NExSS) research coordination network sponsored by NASA’s Science Mission Directorate.

\software{
NumPy \citep{numpy}
SciPy \citep{scipy}
Matplotlib \citep{pyplot}
Astropy \citep{astropy}
EPOS \citep{epos}
emcee \citep{2013PASP..125..306F}
corner \citep{corner}
KeplerPORTs \citep{2017ksci.rept...19B}
}

\ifhasbib
	\bibliography{epos.bbl}
\else	
	\bibliography{papers3,books,software}
\fi

\appendix

\section{Table of Mathematical Symbols}
All mathematical symbols used in this paper are summarized in table \ref{t:sym}.

\begin{table}
	\title{Parameters}
   	\begin{tabular}{lll}
   	\hline\hline
	Parameter & Unit		& Description \\
	\hline
	$R$		& $R_\oplus	$ 	& Planet radius \\
	$P$ 	& 	day			& Planet orbital period \\
	$\mathcal{P}$	& 		& Orbital period ratio of adjacent planets \\	
	$i$		& $\degr$			& Orbital inclination \\
	\hline
	$\delta i$ & $\degr$ 	& Planet mutual inclination \\
	$\Delta i$ & $\degr$ 	& Mode of mutual inclination distribution \\ 			
	$\Omega$ & rad			& Longitude of ascending node \\ 
	\hline
	$f$ 	&				& Planet probability density function \\
	$g$ 	&				& Planetary system probability density function \\
	\hline
	$\chi$	& 				& Continuous random variable \\
	\fdet 	&				& Survey completeness \\
	\fgeo 	& 				& Geometric transit probability \\
	\fsnr 	&				& Detection efficiency \\
	\fvet 	&				& Vetting efficiency \\
	\fiso 	&				& Fraction of isotropic systems \\
		\hline
	$\eta$	& 				& Average number of planets per star \\
	$\eta_s$ 	& 			& Fraction of stars with planets \\
	$m$ 	&				& Planets per system (multiplicity) \\
	ID		& 				& Host star identifier \\
		\hline
	\{\}	& 				& Ensemble of parameters\\
	n 		&				& Sample size \\
	N 		&				& Number of detected planets \\
		\hline
	$s$		& 				& Subscript for planetary systems\\
	$p$		& 				& Subscript for planets\\
	$t$ 	& 				& Subscript for transiting planets \\
	$d$ 	& 				& Subscript for detectable planets \\
	$\text{in}$	& 				& Subscript for innermost planet \\
	$k$		& 				& Index for k-th planet in system \\
	$a,b$ 	& 				& Power law indices \\
	$c$ 	& 				& Normalization factor \\
	$D,\sigma$	&			& Dimensionless spacing parameters \\
   	\hline\hline
   	\end{tabular}
   	\caption{List of mathematical symbols used in this paper.}
	\label{t:sym}
\end{table}

\begin{figure*}
	% from DR25/disposition.py
	% cp ~/DR25/png/*.dwarfs.39975.png  png/
    \includegraphics[width=0.54\linewidth]{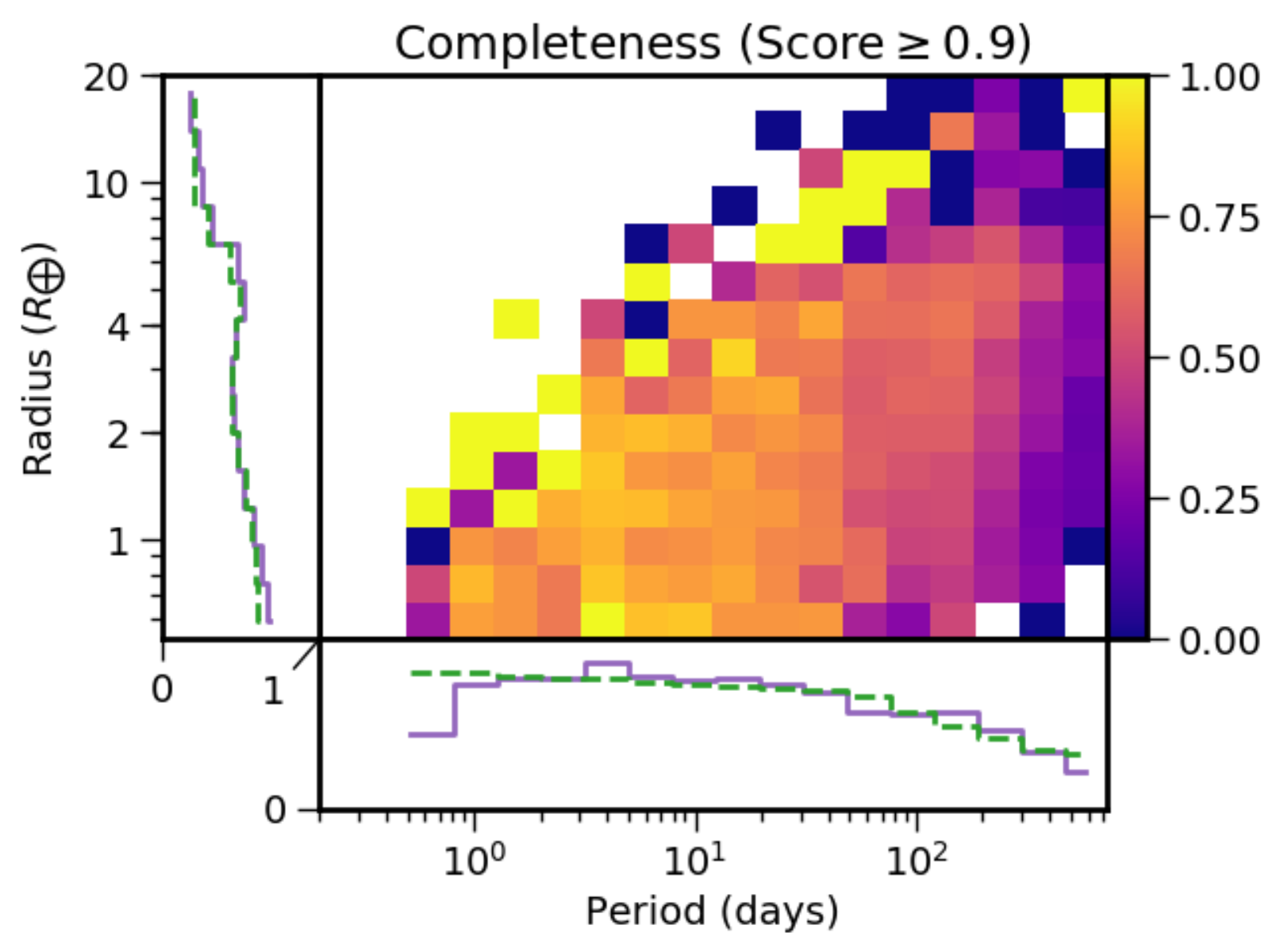}
    \includegraphics[width=0.4\linewidth]{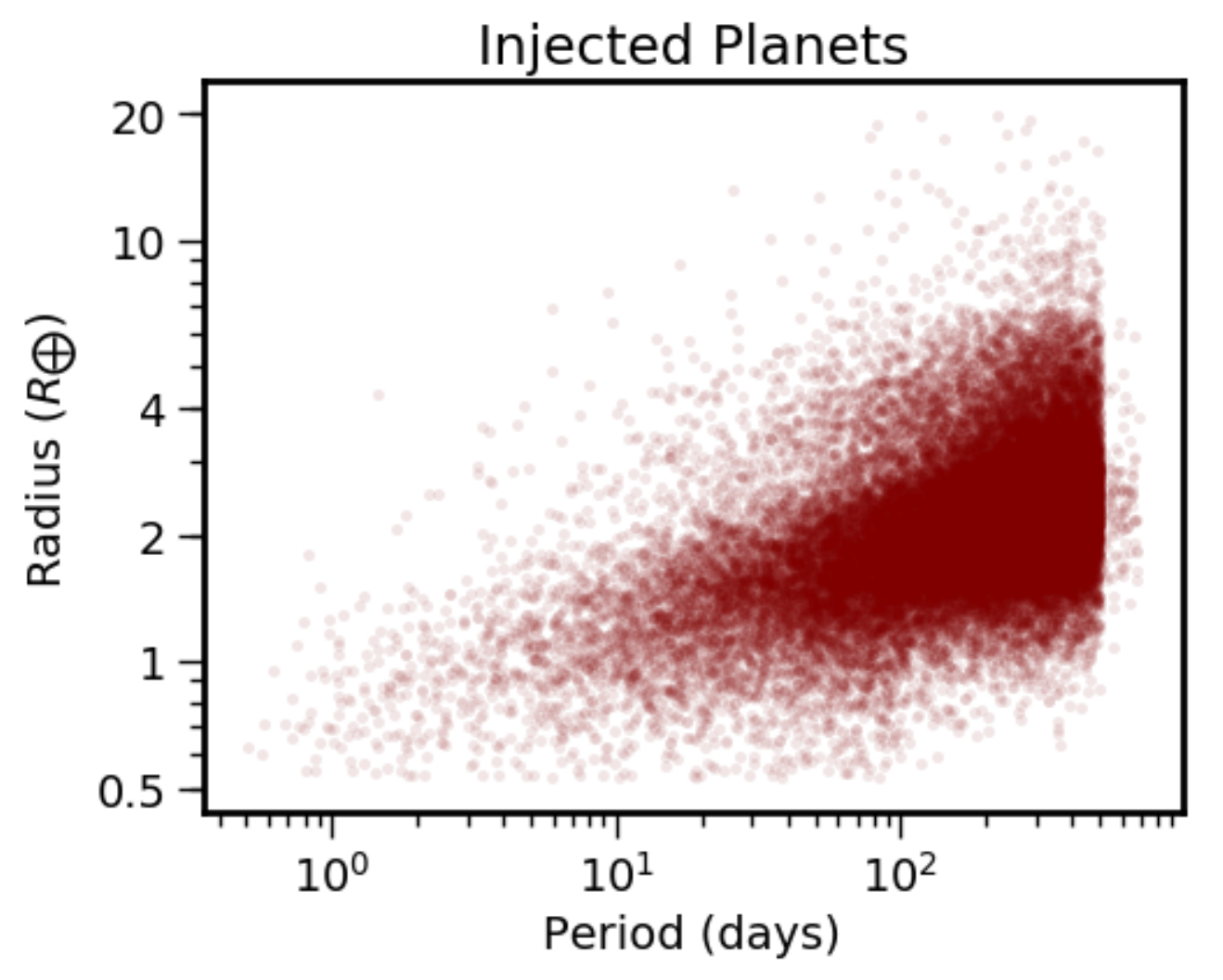}\\
    \includegraphics[width=0.49\linewidth]{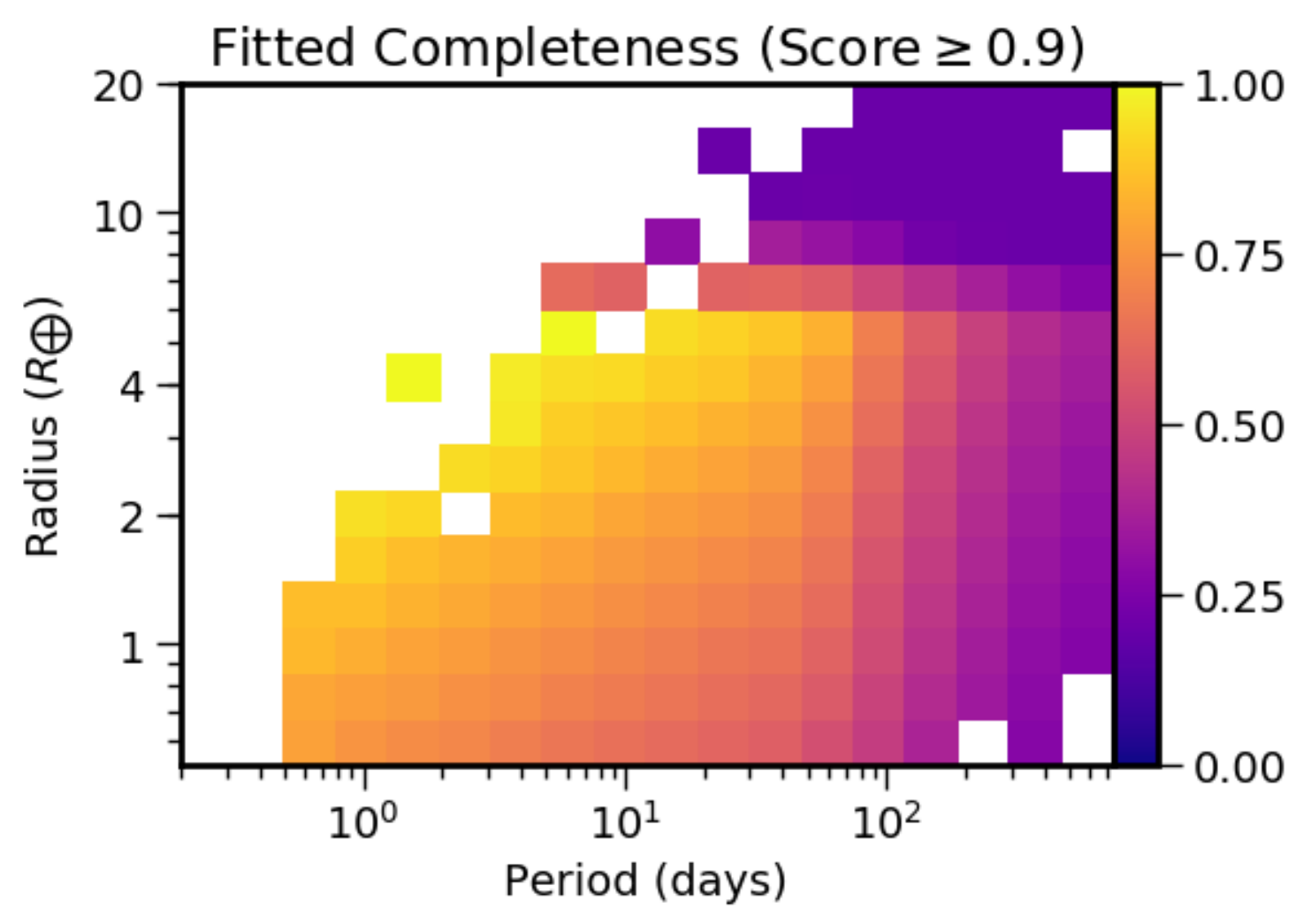}
    \includegraphics[width=0.49\linewidth]{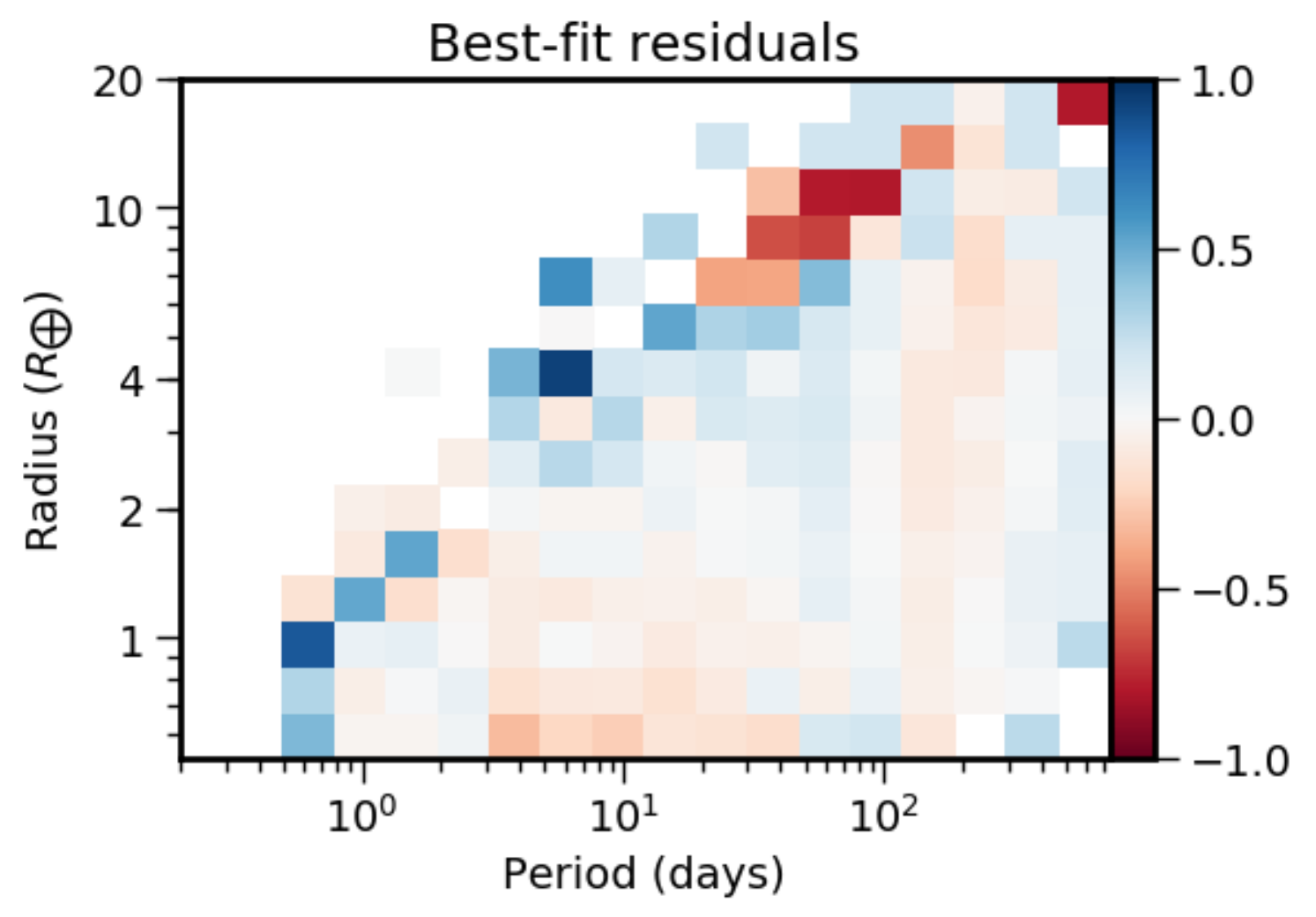}
    \caption{The top left panel show the vetting efficiency per grid cell for a Robovetter score larger than $0.9$. The side panels show marginalized detection efficiency (solid purple) compared to the best-fit double broken power-law model (green dashed). 
    The top right panel shows the locations of the planet injections from \cite{2017ksci.rept...18C}.
    The bottom left panel shows the best-fit detection efficiency evaluated per grid cell.
    The bottom right panel shows the difference between the measured and fitted vetting completeness per grid cell.
    }
    \label{f:vettingdetails}
\end{figure*}

\section{Vetting Completeness}\label{s:vet}
The \kepler Robovetter \citep{2016ApJS..224...12C,2017ksci.rept....1C} was designed to discriminate between real transiting events and instrumental or astrophysical signals that mimic that of a transiting planet, a process called ``vetting''. For planets with few transits and a low signal-to-noise ratio, vetting is a trade-off between detecting planets (completeness) or removing false positives (reliability). A planet candidate sample with high completeness has a low reliability, and vice versa. The Robovetter was calibrated by evaluating the ability to identify the signals of a number of simulated transiting events to better quantify the efficiency and confidence with which planets can be detected. 

Each planet candidate is assigned a disposition score between $0$ and $1$, which can be used to create a more reliable sample (fewer false positives classified as planet candidates) at the the costs of having a less complete sample (more planets classified as false positives), see \cite{2018ApJS..235...38T} for details. For \epos we opt to use a planet sample with high reliability by using a disposition score cut of $0.9$, reducing the sample completeness which, however, can be quantified and corrected for as we discuss below. The score cut of $0.9$ was chosen to eliminate the peak of false positives in the habitable zone that coincides with the orbital period of the spacecraft. 

An empirical procedure for estimating the vetting completeness is described in \cite{2018ApJS..235...38T}, based on injection of simulated transits into the \kepler data \citep{2017ksci.rept...18C} and analyzed by the Robovetter \citep{2017ksci.rept....1C}. We downloaded the Robovetter results for simulated planet injections (group 1) from the \kepler simulated data page\footnote{\url{https://exoplanetarchive.ipac.caltech.edu/docs/KeplerSimulated.html}}. We removed all injections not on main-sequence stars according to the prescription in \cite{2016ApJS..224....2H}.
We only considered injections inside the orbital period and radius range where we use \epos ($P=[0.5,730]$ days and $R=[0.3,20] ~R_\oplus$). There are $44981$ injections, of which $77\%$ ($34558$) are classified as planet candidates and $43\%$ ($19247$) have a disposition score $>0.9$.

The top left panel of Figure \ref{f:vettingdetails} shows the vetting completeness, the fraction of injections classified as planet candidates with a score larger than $0.9$, binned to the same grid as the survey detection efficiency. Planet injections are not equally distributed across this parameter range, with the majority of injections around sub-Neptunes at long orbital periods, as shown in the top right panel.
The vetting completeness is $\sim 75\%$ at orbital periods less than roughly $50$ days and decreases below $25\%$ at the longest orbital periods. The completeness also drops below $25\%$ for giant planets, though this is most likely an artifact of those planets only being injected around the most noisy stars, see \cite{2018ApJS..235...38T}, and does not reflect a real decrease in detectability of giant planets in the \kepler pipeline.

We parametrize the vetting completeness to obtain a smooth distribution of radius and orbital period that covers all of the regions we are simulating with \epos. We use a functional form for the vetting completenss, $\fvet$, that is a double broken power-law:
\begin{equation}
\fvet= c ~f_{\text{vet},P}(P) ~f_{\text{vet},R}(R)
\end{equation}
with
\begin{equation}
f_{\text{vet},P}(P)= 
\begin{cases}
    (P/P_{\rm break})^{a_P},& \text{if } P < P_{\rm break}\\
    (P/P_{\rm break})^{b_P},& \text{otherwise}
\end{cases}
\end{equation}
and
\begin{equation}
f_{\text{vet},P}(R)= 
\begin{cases}
    (R/R_{\rm break})^{a_R},& \text{if } R < R_{\rm break}\\
    (R/R_{\rm break})^{b_R},& \text{otherwise.}
\end{cases}
\end{equation}
We find the best-fit parameters using \texttt{scipy.optimize.curve\_fit}, where we minimize the difference between the measured and parametrized vetting completeness over all grid cells.
The best-fit solution is shown in the bottom left panel of Figure \ref{f:vettingdetails} and the bottom right panel shows the residuals per grid cell. 
The best-fit parameters are 
$c=0.88 \pm 0.08$, 
$P_{\rm break}= 53 \pm 8$ days,
$a_P= -0.07 \pm 0.03$,
$b_P= -0.39 \pm 0.02$,
$R_{\rm break}=  5.7 \pm 2.1 ~R_\oplus$,
$a_R= 0.19 \pm 0.03$, and
$b_R= -2.7 \pm 6.2$.
The break in the power-law with orbital period corresponds to the decreased detection efficiency at long orbital periods as documented in \cite{2018ApJS..235...38T,2017ksci.rept....1C}. The break in the power-law of planet radius corresponds to a decreased detection efficiency for giant planets. However, as mentioned above, this is likely an artifact of the way the injection tests were defined, and there is no reason to assume that the low vetting efficiency for giant planets is a real feature \citep{2018ApJS..235...38T}. Therefore, we do no include the power-law index radius break in \epos. Instead, we assume the planet radius dependence of the detection is described by the power-law in planet radius that was fitted to the small planets. This yields a vetting efficiency of
\begin{equation}
\fvet(R,P)= 0.63 ~R^{0.19}
\begin{cases}
    (P/P_{\rm break})^{-0.07},& \text{if } P < P_{\rm break}\\
    (P/P_{\rm break})^{-0.39},& \text{otherwise}
\end{cases}
\end{equation}
with $P_{\rm break}= 53$, which is shown in Figure \ref{f:eff}.

\section{Posterior Distributions}
\begin{figure*}
% cp ~/EOStool/png/dr25_vetting/mcmc/chain.png png/
    \includegraphics[width=\linewidth]{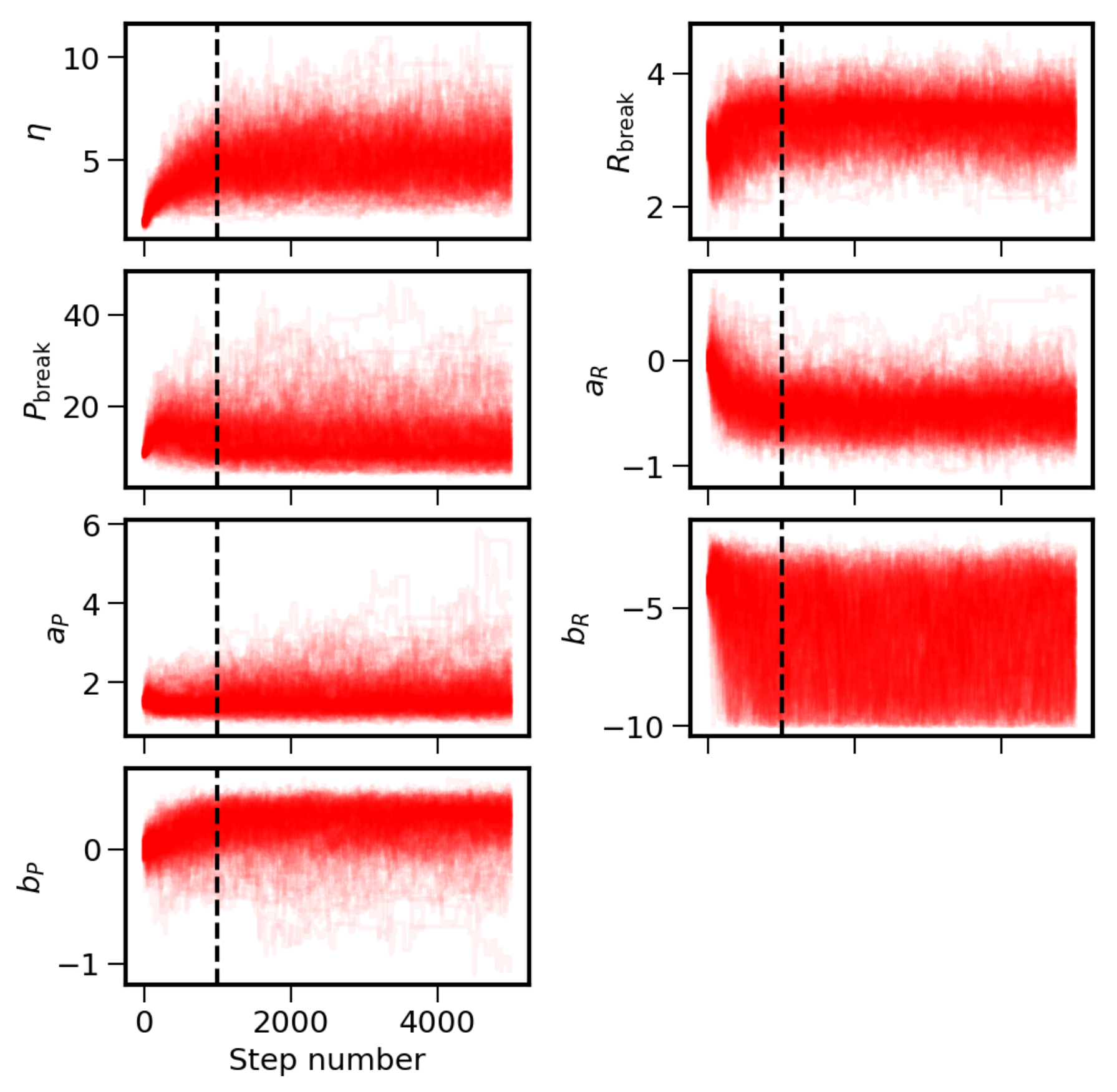}
    \caption{Position in 7-dimensional parameter space of 200 walkers in the MCMC chain. The dashed line indicates the end of the burn-in phase.}
    \label{f:chain}
\end{figure*}

\begin{figure*}
% cp ~/EOStool/png/dr25_vetting/mcmc/triangle.png png/
    \includegraphics[width=\linewidth]{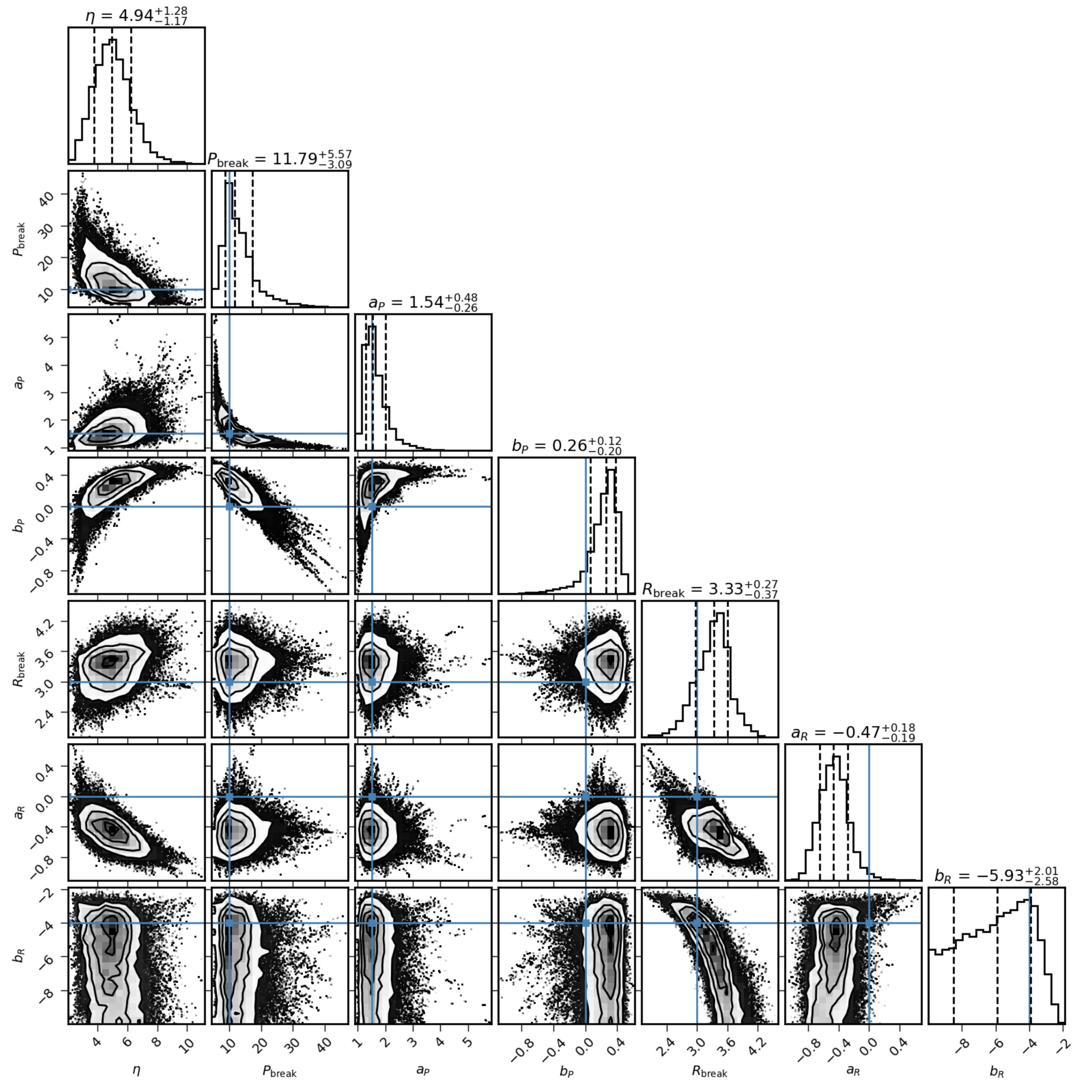}
    \caption{Corner plots for the parametric distribution, generated using open-source \python package \texttt{corner}.
    Blue lines indicate the initial guess that was based on previous studies.}
    \label{f:corner}
\end{figure*}

\begin{figure*}
	% cp ~/EOStool/png/dr25_vetting_multi/mcmc/triangle.png png/multi_corner.png
    \includegraphics[width=\linewidth]{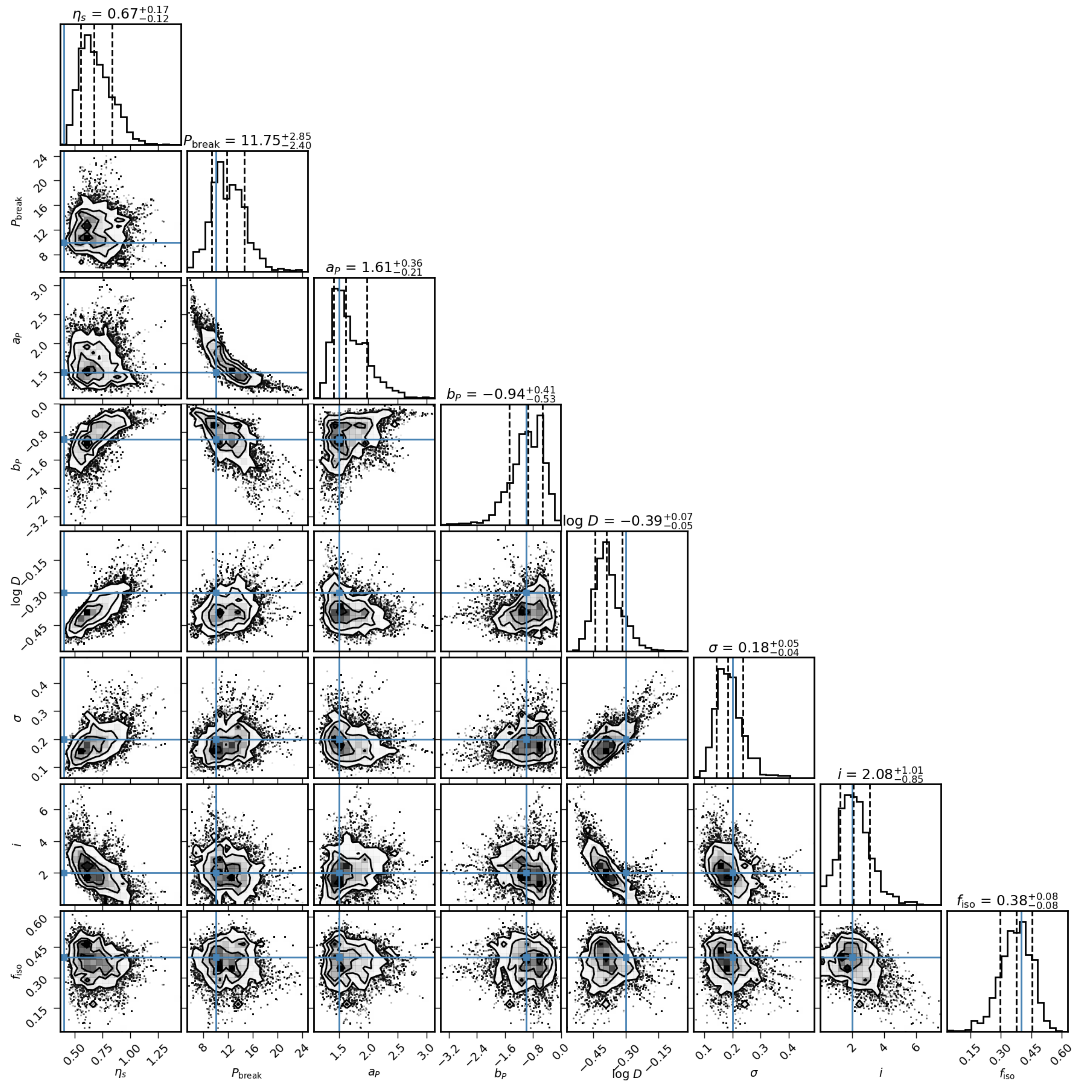}
    \caption{Corner plots for the multi-planet parametric distribution, generated using open-source \python package \texttt{corner}.
    Blue lines indicate the initial guess. }
    \label{f:multi:corner}
\end{figure*}

The MCMC chain and corner plots for occurrence rate mode are shown in Fig. \ref{f:chain} and Fig. \ref{f:corner}. The corner plots for multi-planet mode are shown in Fig. \ref{f:multi:corner}

\end{document}